\documentstyle[psfig]{mn}
\newif\ifAMStwofonts
\newcommand{\gapp}{\mbox{\raisebox{-0.3em}{$\stackrel{\textstyle >}{\sim}$}}}

\title[A multifrequency study of 3C46 and 3C452]{A multifrequency study of the large radio galaxies 3C46 and 3C452} 
\author[S. Nandi et al.]
       {S. Nandi,$^{1}$$\thanks{E-mail: sumana@aries.res.in(SN), akash@aries.res.in(AP), spal@cyllene.uwa.edu.au(SP), 
        chiranjib.konar@gmail.com(CK),  djs@ncra.tifr.res.in, Dhruba.Saikia@csiro.au(DJS), msingh@aries.res.in(MS)}$ 
	A. Pirya,$^{1}$ S.Pal,$^{2,3}$ C. Konar,$^{4}$ D.J. Saikia,$^{2,3,5}$ and M. Singh$^{1}$ \\
$^{1}$ Aryabhatta Research Institute of Observational Sciences (ARIES),  Manora Peak, Nainital 263 129, India \\
$^{2}$ International Centre for Radio Astronomy Research, The University of Western Australia, Crawley, WA 6009, Australia \\
$^{3}$ National Centre for Radio Astrophysics, TIFR, Pune University Campus, Post Bag 3, Pune 411 007, India \\
$^{4}$ Indian Institute of Astrophysics, Sarjapur Road, Koramangala, Bangalore 560 034, India \\
$^{5}$ Australia Telescope National Facility, CSIRO, PO Box 76, Epping, NSW 1710, Australia         
}
\date{Accepted.    Received }

\pagerange{\pageref{firstpage}--\pageref{lastpage}}
\pubyear{  }

\begin{document}

\maketitle

\label{firstpage}

\begin{abstract}
We present low-frequency observations starting from $\sim$150 MHz with the Giant Metrewave Radio 
Telescope (GMRT), and high-frequency observations with the Very Large Array (VLA) of 
two large radio galaxies 3C46 and 3C452. These observations were made with the objectives
of estimating their spectral ages and examining any evidence of diffuse extended emission at
low radio frequencies due to an earlier cycle of activity. While no evidence of extended
emission due to an earlier cycle of activity has been found, the spectral ages have been estimated 
to be { $\sim$15 and 27 Myr for the oldest relativistic plasma seen in the regions close to 
the cores for 3C46 and 3C452 respectively.}
The spectra in the vicinity of the hotspots are consistent with a straight spectrum
with injection spectral indices of $\sim$1.0 and 0.78 respectively, somewhat steeper than
theoretical expectations.
\end{abstract}


\section{Introduction}
Large double-lobed radio galaxies, the largest of which are the giant radio sources 
(GRSs), defined to be those which have a projected linear size
$\gapp$1 Mpc (H$_o$=71 km s$^{-1}$ Mpc$^{-1}$, $\Omega_m$=0.27,
$\Omega_{vac}$=0.73, Spergel et al. 2003), are useful for studying the late stages 
of evolution of radio sources and possible episodic activity in these objects, constraining 
orientation-dependent unified schemes and probing the intergalactic medium at different 
redshifts (e.g. Subrahmanyan \& Saripalli 1993; Subrahmanyan, Saripalli \& Hunstead 1996;
Mack et al. 1998; Ishwara-Chandra \& Saikia 1999;
Kaiser \& Alexander 1999; Blundell, Rawlings \& Willott 1999 and references therein; 
Schoenmakers et al. 2000, 2001; Singal, Konar \& Saikia 2004).
In addition, large radio sources are useful for studying the 
effects of electron energy loss in the lobe plasma 
due to inverse-Compton scattering with the Cosmic Microwave Background Radiation (CMBR) photons 
at different redshifts (e.g. Konar et al. 2004), make independent estimates of the magnetic
field from the inverse-Compton scattered X-ray flux density from the lobes 
(e.g. Croston et al. 2005; Konar et al. 2009) and spectral as well as dynamical ageing analyses 
to understand the evolution of the sources (e.g. Konar et al. 2006, 2008; 
Jamrozy et al. 2008; Machalski, Jamrozy \& Saikia 2009). 

%%%%%%%%%%%%%%%%%%%%%%%%%%%%%%%%%%%%%%%%%%%%%%%%%%%%%%%%%%%%%%%%%%%%%%%%%%%%%%%
\begin{figure*}
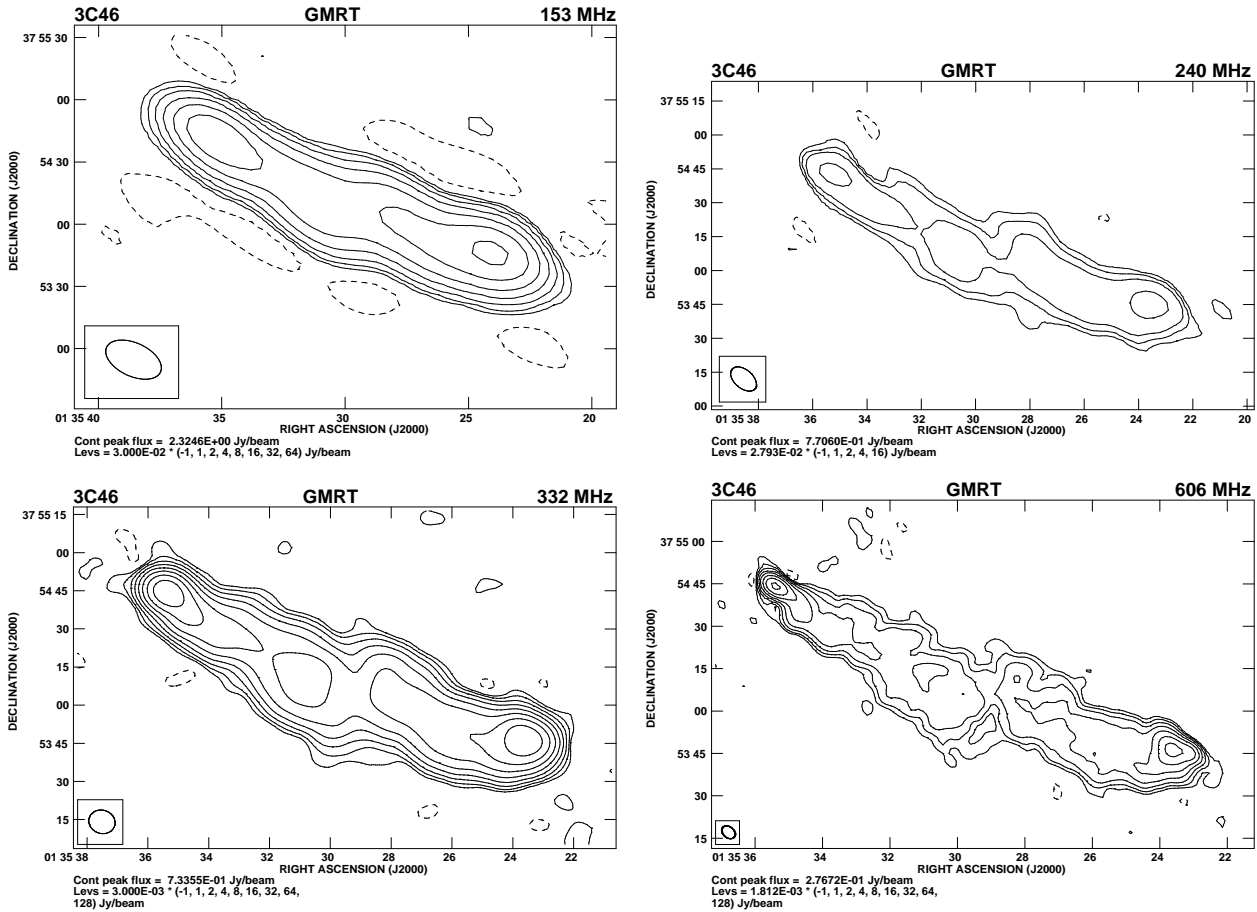

\vbox{
  \hbox{
   \psfig{file=3C46_153NEW.PS,width=3.3in,angle=-90}
   \psfig{file=3C46_240.PS,width=3.3in,angle=-90}
       }
  \hbox{
   \psfig{file=3C46_325.PS,width=3.3in,angle=-90}
   \psfig{file=3C46_606.PS,width=3.3in,angle=-90}
       }
    }
\caption[]{GMRT low-frequency images of 3C46 at 153, 240, 332 and 606 MHz, 
and VLA higher-frequency images at 1465, 4841 and 8460 MHz. 
In this figure as well as in all the other images of the sources, 
the peak brightness and the contour levels are given below each image. 
In all the images the restoring beam is indicated by an ellipse. 
          }
\end{figure*}

%%%%%%%%%%%%%%%%%%%%%%%%%%%%%%%%%%%%%%%%%%%%%%%%%%%%%%%%%%%%%%%%%%%%%%%%%%%%%%%

We have selected two large radio galaxies with prominent bridges, 3C46 and 3C452, for making 
detailed radio images
with the Giant Metrewave Radio Telescope (GMRT) at low frequencies going down to $\sim$150 MHz.
We have also used archival Very Large Array (VLA) data to make higher-frequency images of these
two sources. The twin objectives of this study were to look for diffuse emission at low
frequencies from an earlier cycle of activity and estimate the spectral ages of the lobes
from data over a large frequency range. Combining the lowest-frequency available data with
high-frequency data gives the most reliable estimates of the injection spectral indices 
($\alpha_{\rm{inj}}$), and also spectral ages from the break frequency. { For examining
emission from an earlier cycle of activity, it is relevant to note that evidence of episodic 
activity is seen usually in large radio sources (e.g. Schoenmakers et al. 2000; Saikia, Konar 
\& Kulkarni 2006, and references therein) but not in the small sources even at low 
radio frequencies (Sirothia et al. 2009).}

The radio galaxy 3C46 (J0135+3754), which has a very close companion galaxy (de Vries et al. 1998), 
is at a redshift of 0.4373 {(Smith \& Spinrad 1980)} and has a largest angular size
of $\sim$150 arcsec which corresponds to 846 kpc, and a total radio luminosity 
of log P$_{1.4 {\rm GHz}}$ (W Hz$^{-1}$) = 27.01 (Konar et al. 2004). 
VLA B- and C-array images at L-band (Gregorini et al. 1988; Vigotti et al. 1989) and the
D-array image at C-band (Konar et al. 2004) show the extended lobes of emission with 
an edge-brightened structure. de Koff et al. (2000) report evidence of a dust lane
lying across the nucleus with the radio axis being nearly perpendicular to it.

3C452 (J2245+3941) is at a redshift of 0.0811 {(Schmidt 1965)}, and has a moderately 
strong core with a flux density
of 130 mJy at 5 GHz (Riley \& Pooley 1975), and two edge-brightened lobes separated by
252 arcsec which corresponds to 381 kpc. The total radio luminosity at 1.4 GHz is 
log P$_{1.4 {\rm GHz}}$ (W Hz$^{-1}$) = 26.53.
There is a suggestion of a faint dust lane near the nucleus leading to the dumbell
shape (de Koff et al. 2000). The axis defined by the radio lobes and the core is
perpendicular to the proposed dust lane.
Very Long Baseline Interferometric (VLBI) observations of the nuclear source show a 
reasonably symmetric structure with the central peak being identified with the core. The jet
symmetry and the prominence of the possible core suggests that the source is
oriented at an angle larger than $\sim$60$^\circ$ to the line of sight ({ Giovannini} et al. 2001).
Gupta \& Saikia (2006) have reported the discovery of H{\sc i} absorption towards the radio core 
of 3C452 from GMRT observations of the source.

%%%%%%%%%%%%%%%%%%%%%%%%%%%%%%%%%%%%%%%%%%%%%%%%%%%%%%%%%%%%%%%%%
\begin{table}
\caption{ Observing log }
\begin{tabular}{l c c c c c}

\hline
Source    & Teles-     & Array  & Obs. & {    Phase }   & Obs.        \\
          & cope       & Conf.  & Freq.& {    Calib.}   & Date        \\
          &            &        & MHz  &               &             \\
  (1)     &  (2)       & (3)    &  (4) & {    (5)  }    & (6)         \\
\hline
3C46      & GMRT       &        & 153  & {     3C48}    & 2007 Dec 07 \\
3C46      & GMRT       &        & 240  & {     3C48}    & 2007 Jun 09 \\
3C46      & GMRT       &        & 332  & {     3C48}    & 2008 Feb 23 \\
3C46      & GMRT       &        & 606  & {     3C48}    & 2007 Jun 09 \\
%-----------------------------------------------------------
3C46      & VLA$^a$    & BnC    & 1465 & {    2254+247} & 2000 Mar 13 \\
3C46      & VLA$^a$    & D      & 4841 & {    2250+143} & 2000 Jul 24 \\
3C46      & VLA$^a$    & D      & 8460 & {    2251+158} & 1998 Jan 24 \\
%-----------------------------------------------------------
3C452     &  GMRT      &        & 153  & {    2350+646} & 2007 Dec 08 \\
3C452     &  GMRT      &        & 240  & {    2350+646} & 2005 May 30 \\
3C452     &  GMRT      &        & 332  & {    2350+646} & 2008 Jan 02 \\
3C452     &  GMRT      &        & 606  & {    2350+646} & 2005 May 30 \\
3C452     &  GMRT      &        & 1314 & {    2202+422} & 2005 Dec 10 \\
%-----------------------------------------------------------
3C452     &  VLA$^a$   &  D     & 4910 & {    2253+417} & 1995 Apr 11 \\
3C452     &  VLA$^a$   &  D     & 8350 & {    2255+420} & 1995 Apr 11 \\
\hline
\end{tabular}

$^a$ archival data from the VLA \\
\end{table}
%%%%%%%%%%%%%%%%%%%%%%%%%%%%%%%%%%%%%%%%%%%%%%%%%%%%%%%%%%%%%%%%%%%

\section{Observations and analyses} 
Both the GMRT and the VLA observations were made 
in the standard fashion, with each target source 
observations interspersed with observations of the phase calibrator. 
The primary 
flux density and bandpass calibrator was 3C48 and/or 3C286 at the different
frequencies, with all flux densities being on the scale of Baars et al. (1977). 
The total observing time on the source is about a few hours for the 
GMRT observations while for the VLA observations the time on source
ranges from a few minutes to $\sim$10 minutes. The low-frequency GMRT data were 
sometimes significantly affected by radio frequency interference,
and these data were flagged. All the data were analysed in the standard 
fashion using the NRAO {\tt AIPS} package. { For the GMRT observations, besides 
flagging bad data, the steps followed include gain calibration of one spectral
channel data, bandpass calibration and channel averaging to obtain the continuum
data base. These were then imaged and CLEANed using multiple facets for the different
low-frequency GMRT observations. All the data were self calibrated 
to produce the final images, which were then corrected for the gain of the primary beam.} 

The observing log for both the GMRT and the VLA observations is given in 
Table 1 which is arranged as follows. Columns 1 and 2 show the name of the 
source and the telescope; column 3 gives the array configuration for the VLA observations;
column 4 shows the frequency of the observations in MHz, { column 5 lists the phase
calibrators used for the different observations}, while 
column 6 lists the dates of the observations. For the VLA observations of 3C452 at 8350 MHz,
there were two sets of observations, one pointed towards each lobe.

%%%%%%%%%%%%%%%%%%%%%%%%%%%%%%%%%%%%%%%%%%%%%%%%%%%%%%%%%%%%%%%%%%%%%%%%%%%%%%%%%%%%%%%

%%%%%%%%%%%%%%%%%%%%%%%%%%%%%%%%%%%%%%%%%%%%%%%%%%%%%%%%%%%%%%%%%%%%%%%%%%%%%%%
\begin{figure}
\vbox{
   \psfig{file=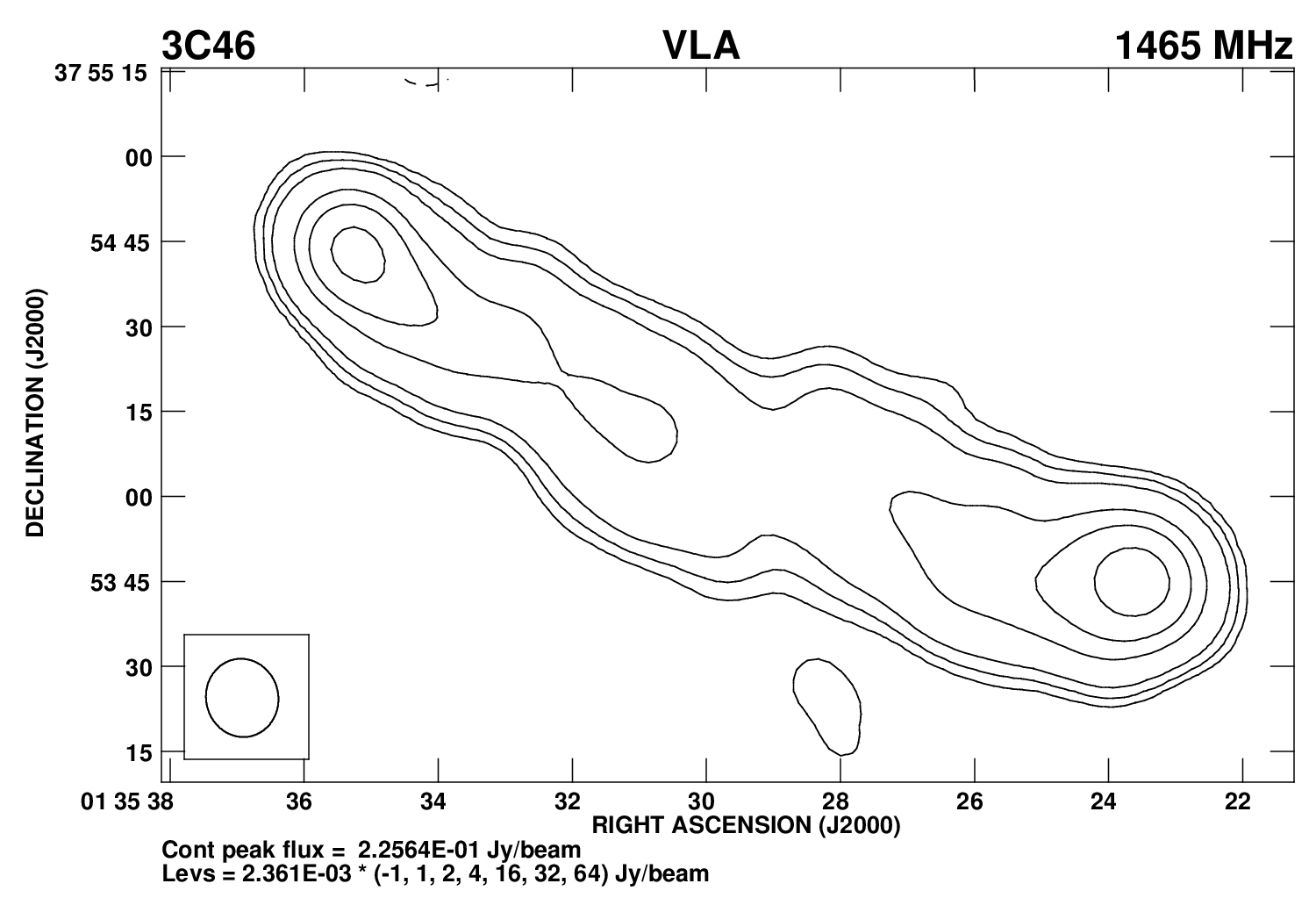,width=3.3in,angle=-90}
   \psfig{file=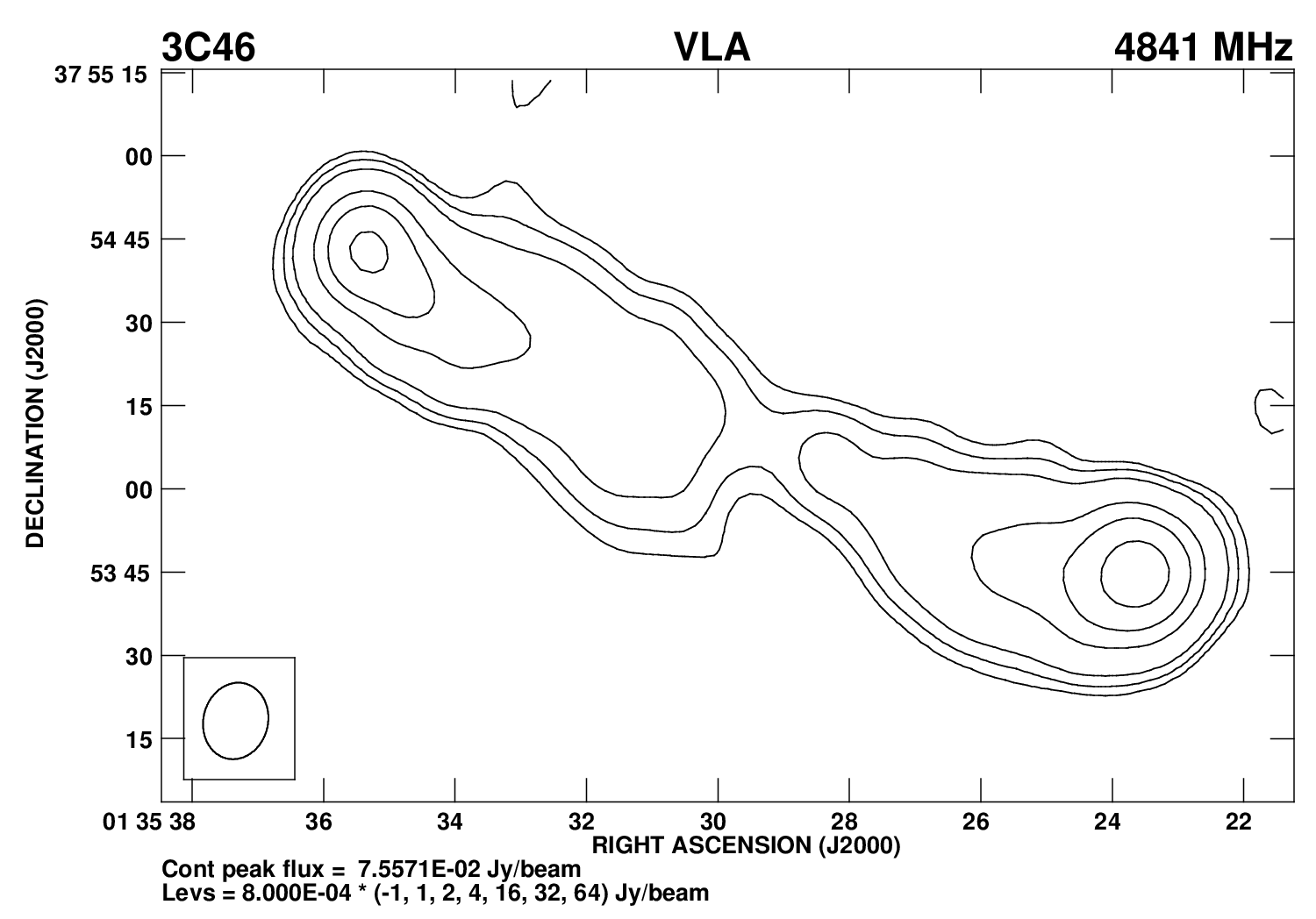,width=3.3in,angle=-90}
   \psfig{file=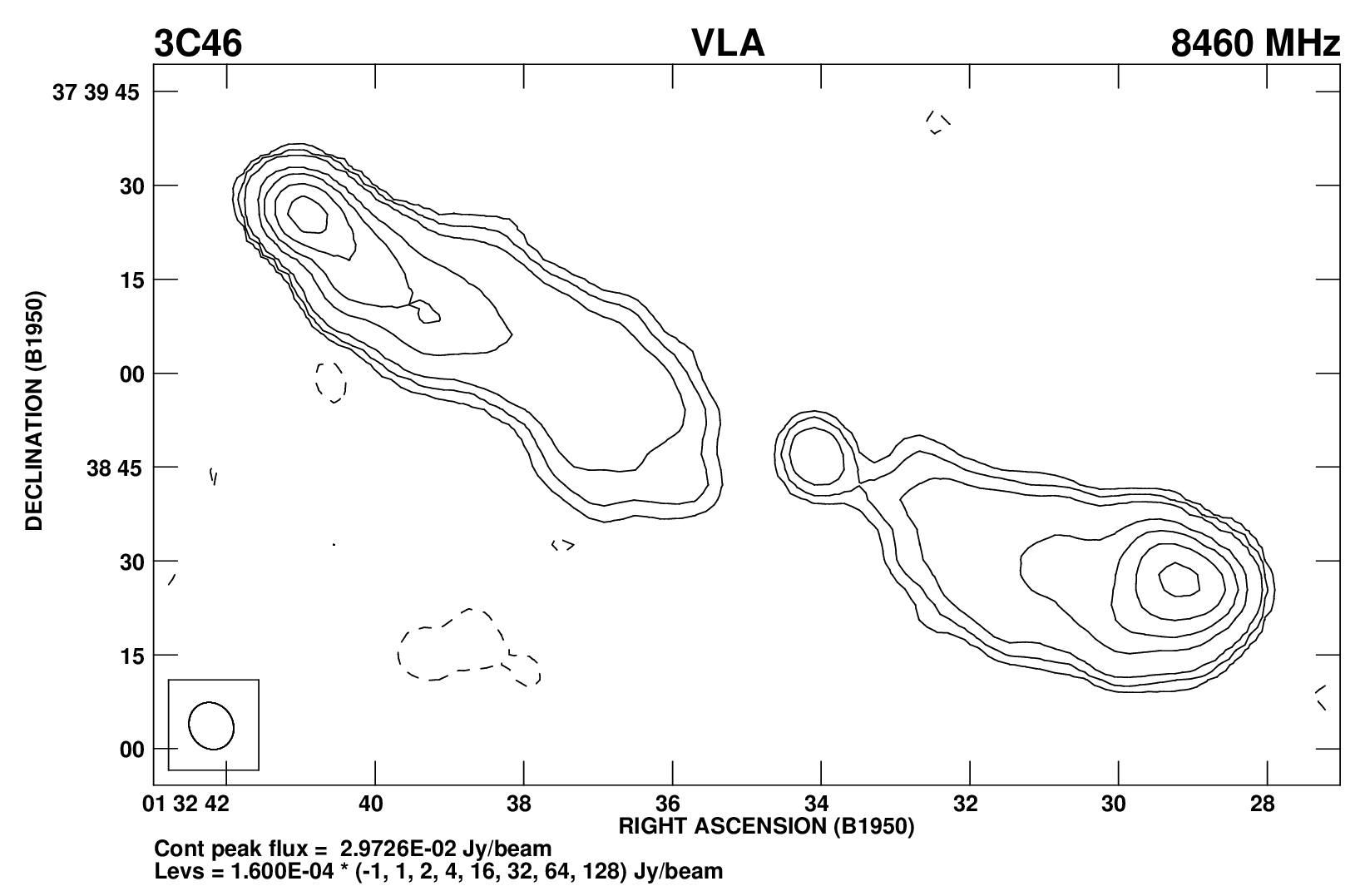,width=3.3in,angle=-90}
    }
\contcaption{}
\end{figure}

%%%%%%%%%%%%%%%%%%%%%%%%%%%%%%%%%%%%%%%%%%%%%%%%%%%%%%%%%%%%%%%%%%%%%%%%%%%%%%%%%%%%%%%%%%%%%%%%%%%%%%%%%%%%%%%%%%%

\begin{figure}
\vbox{
   \psfig{file=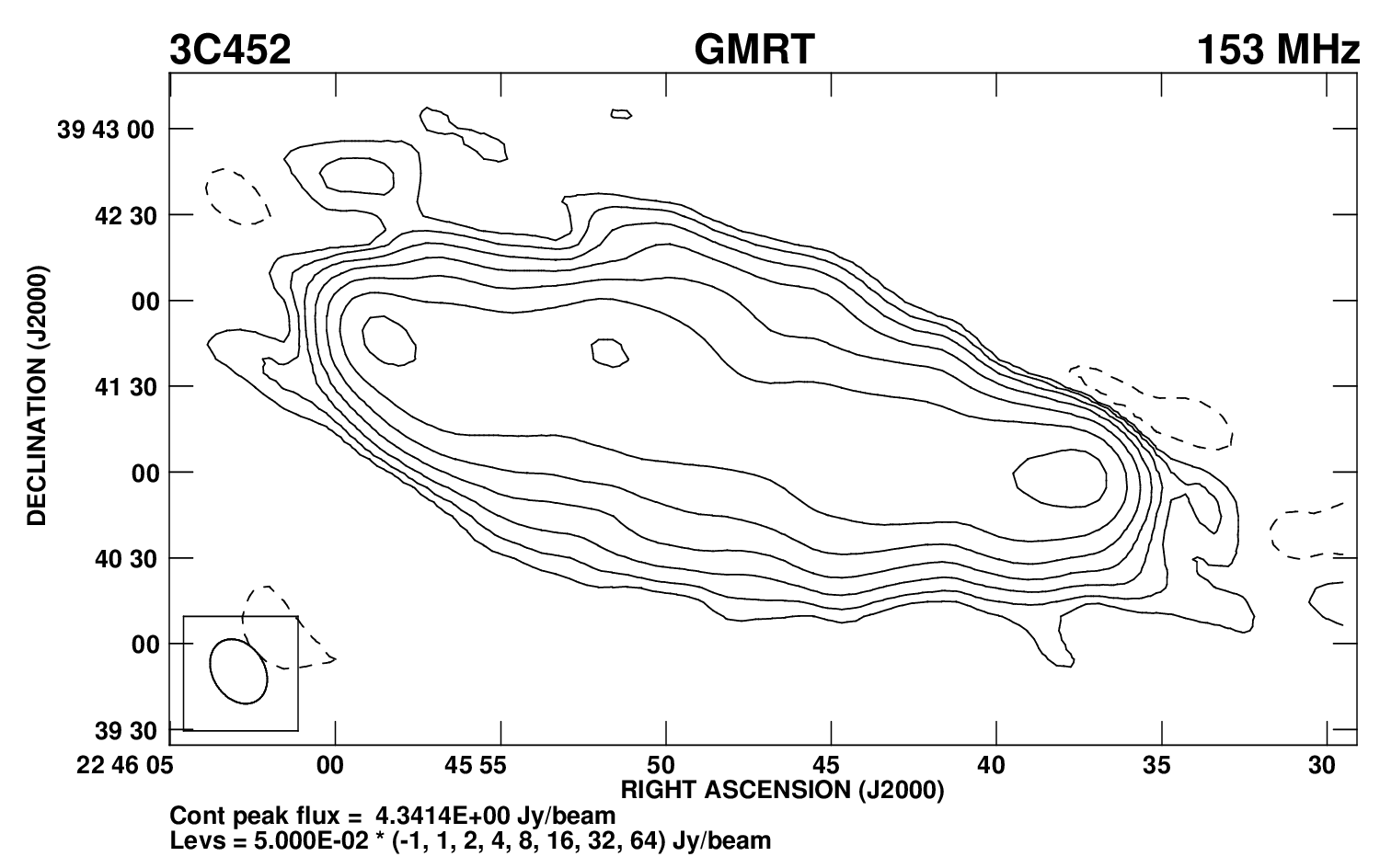,width=3.3in,angle=-90}
   \psfig{file=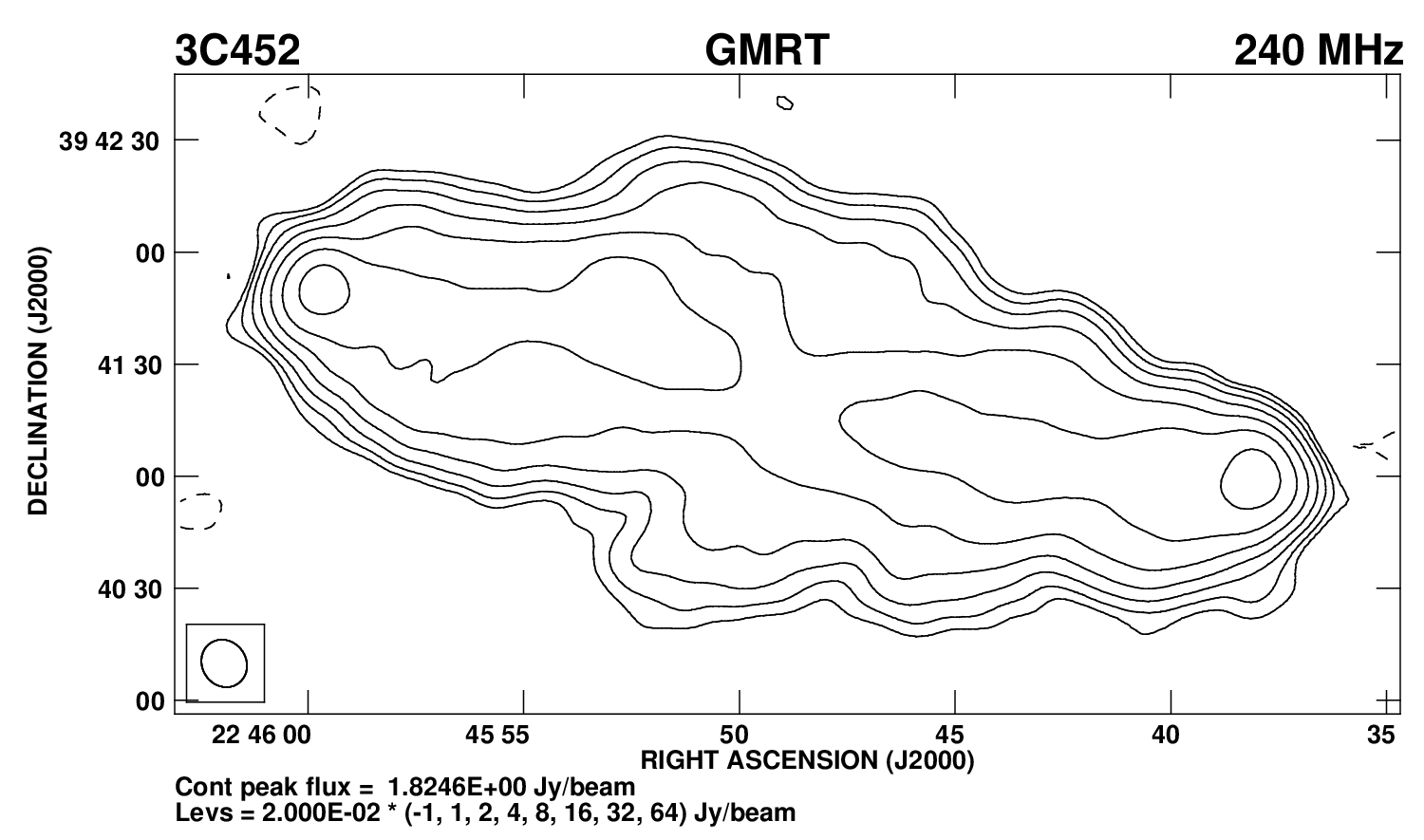,width=3.3in,angle=-90}
   \psfig{file=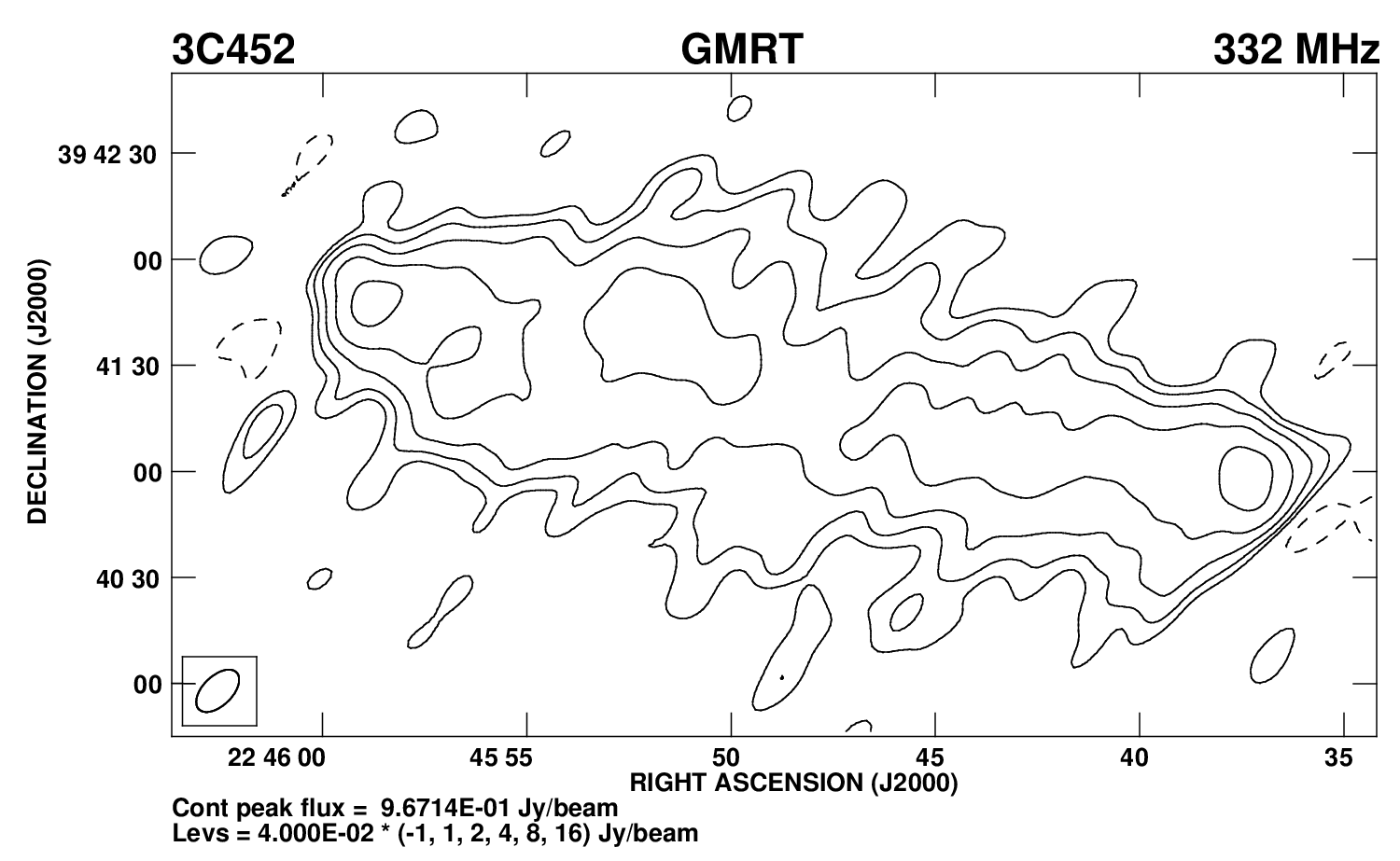,width=3.3in,angle=-90}
   \psfig{file=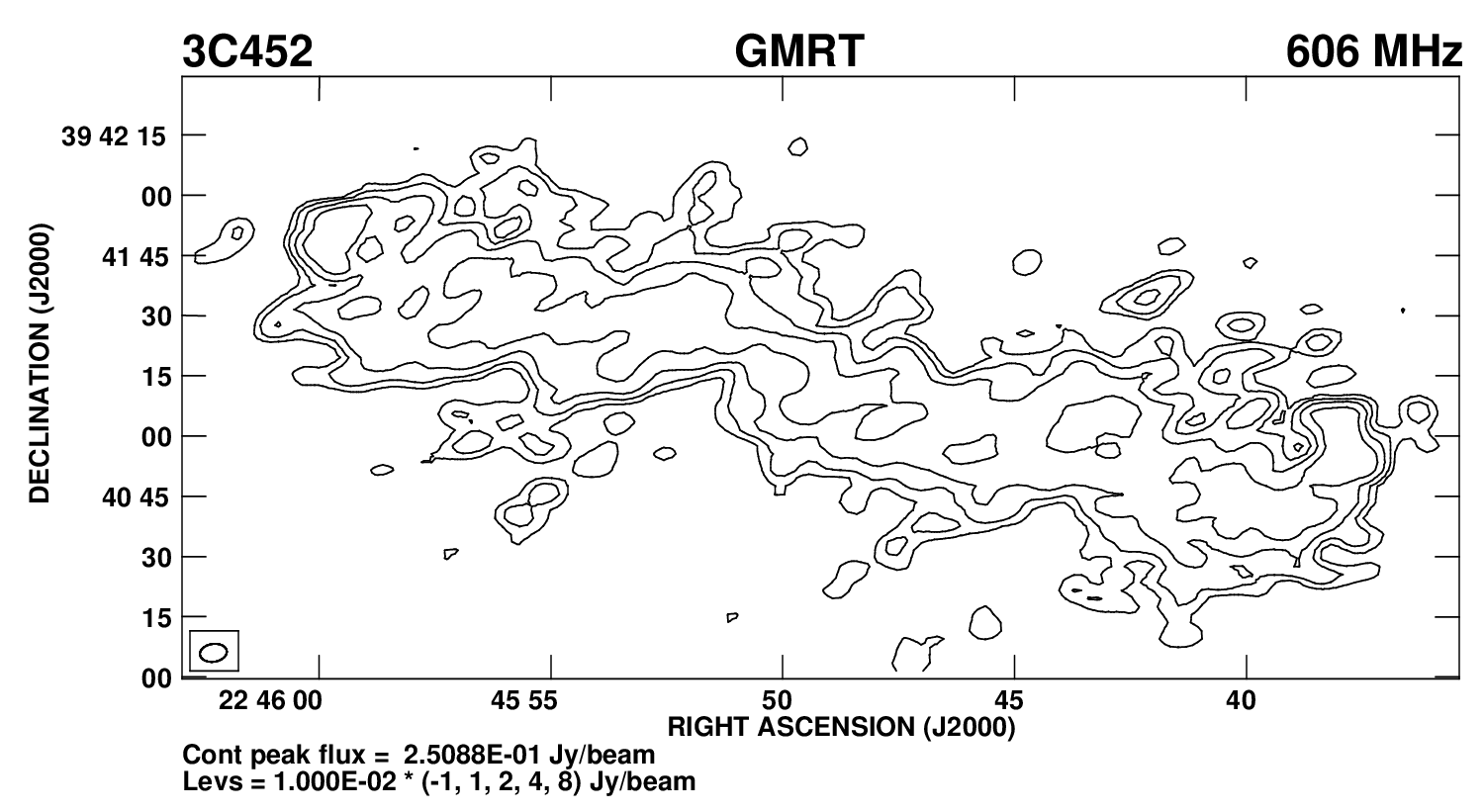,width=3.3in,angle=-90}
     }
\caption[]{GMRT low-frequency images of 3C452 at 153, 240, 332, 606 and 1314 MHz, 
           and VLA higher-frequency images at 4910 and 8350 MHz. 
	   For the images at 8350 MHz, in the case of 3C452W the pointing centre was towards the western hotspot, 
           while in the case of 3C452E the pointing centre was towards the eastern hotspot. 
          }
     \end{figure}

\begin{figure}
\vbox{
   \psfig{file=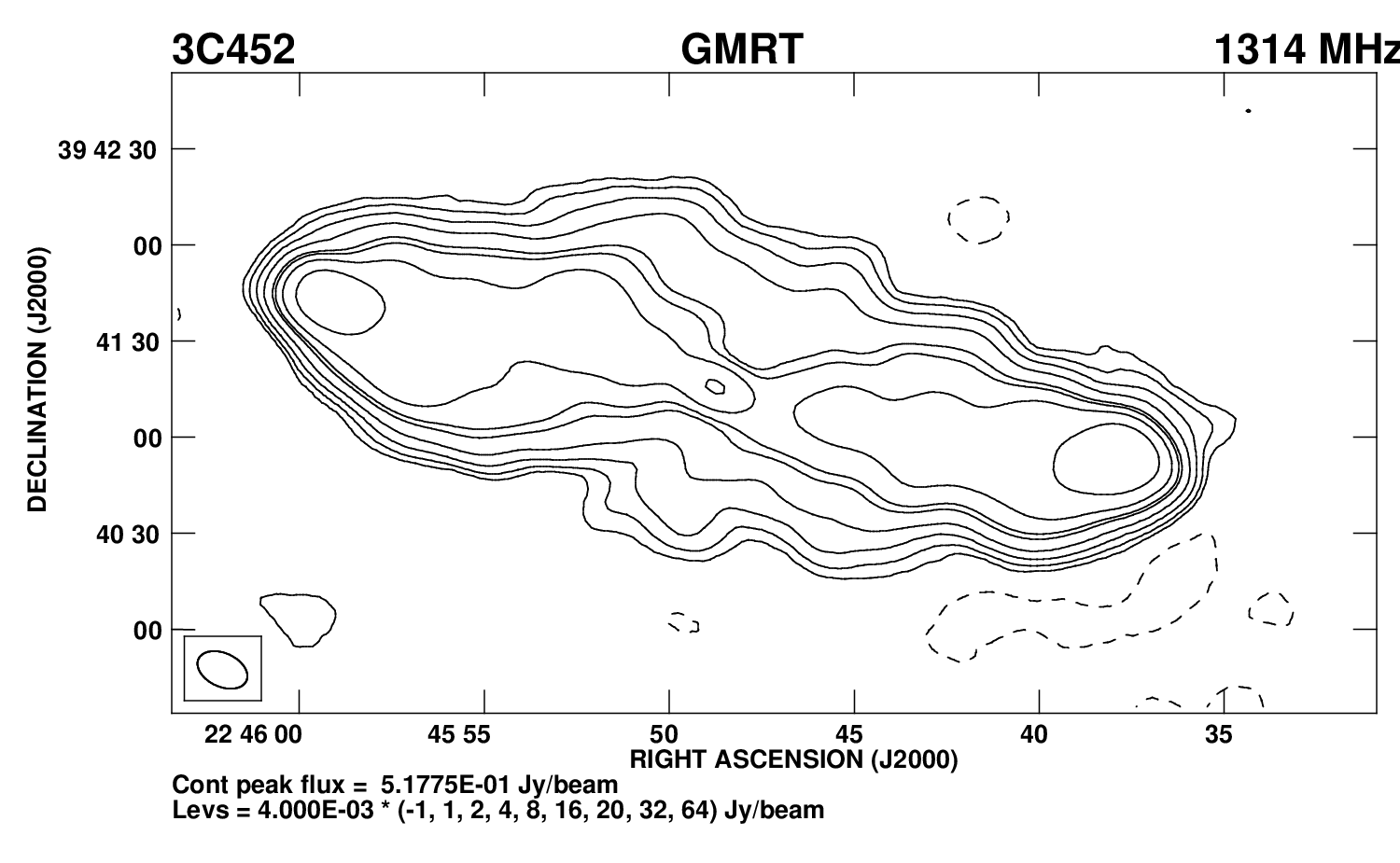,width=3.3in,angle=-90}
   \psfig{file=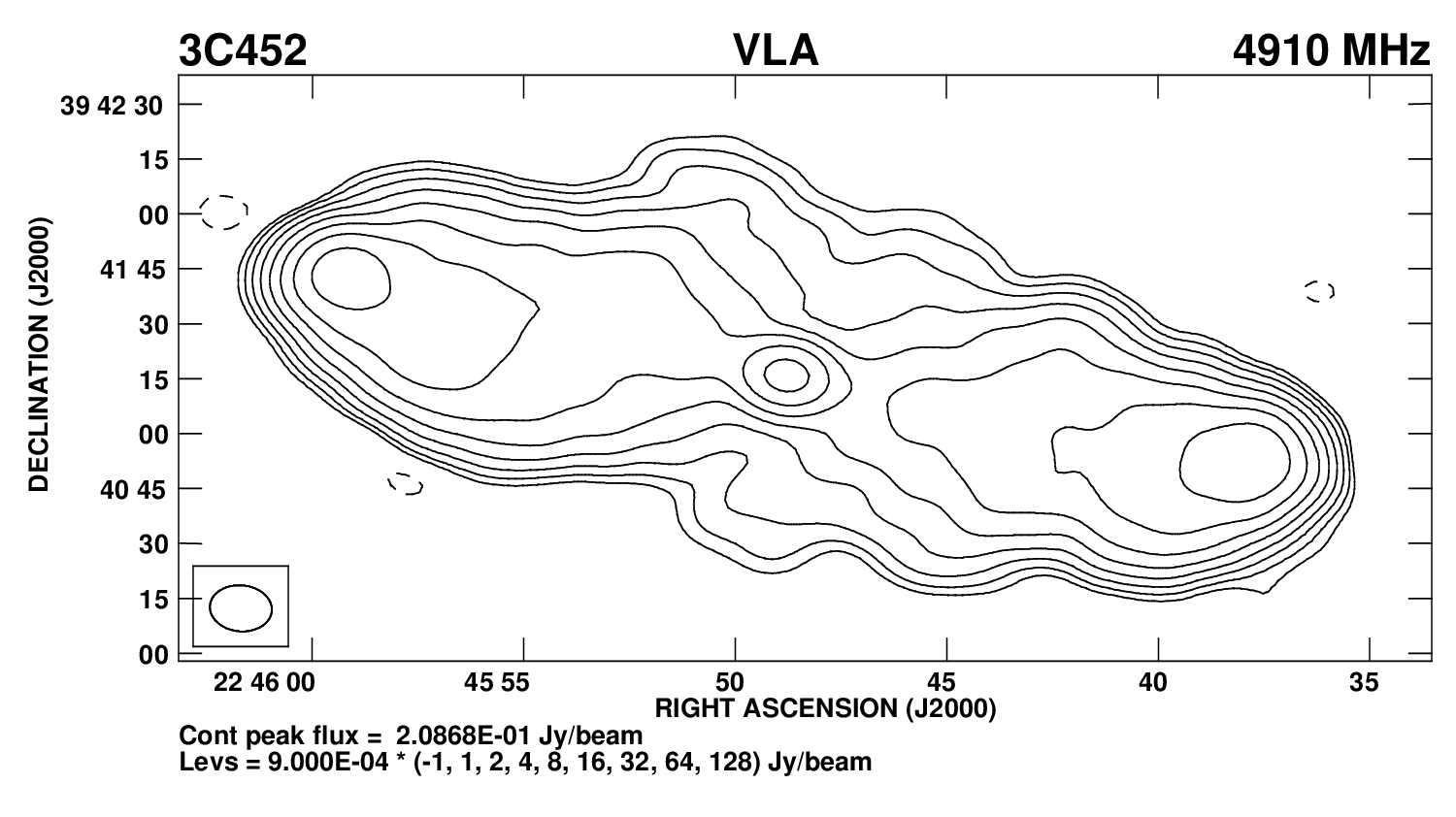,width=3.3in,angle=-90}
   \psfig{file=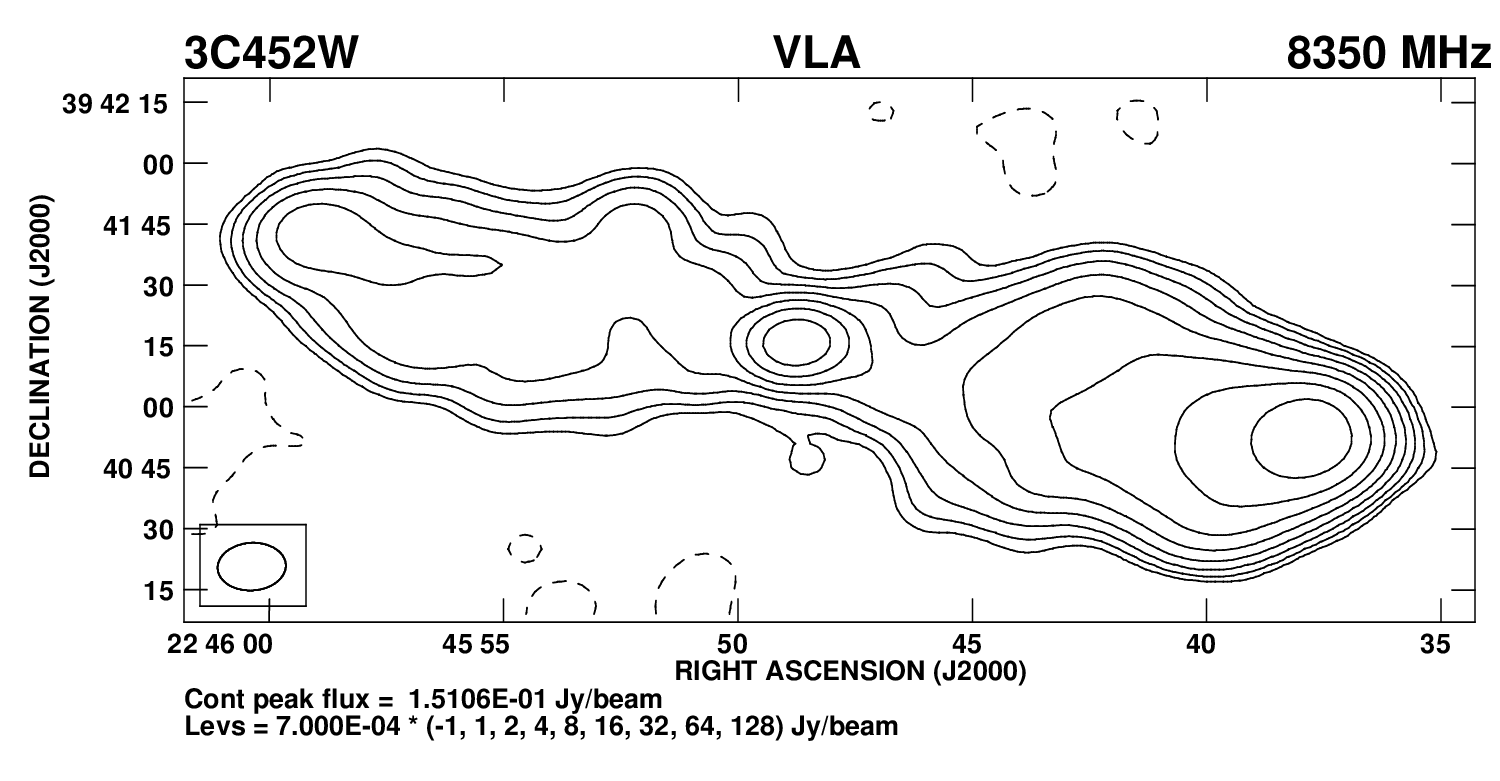,width=3.3in,angle=-90}
   \psfig{file=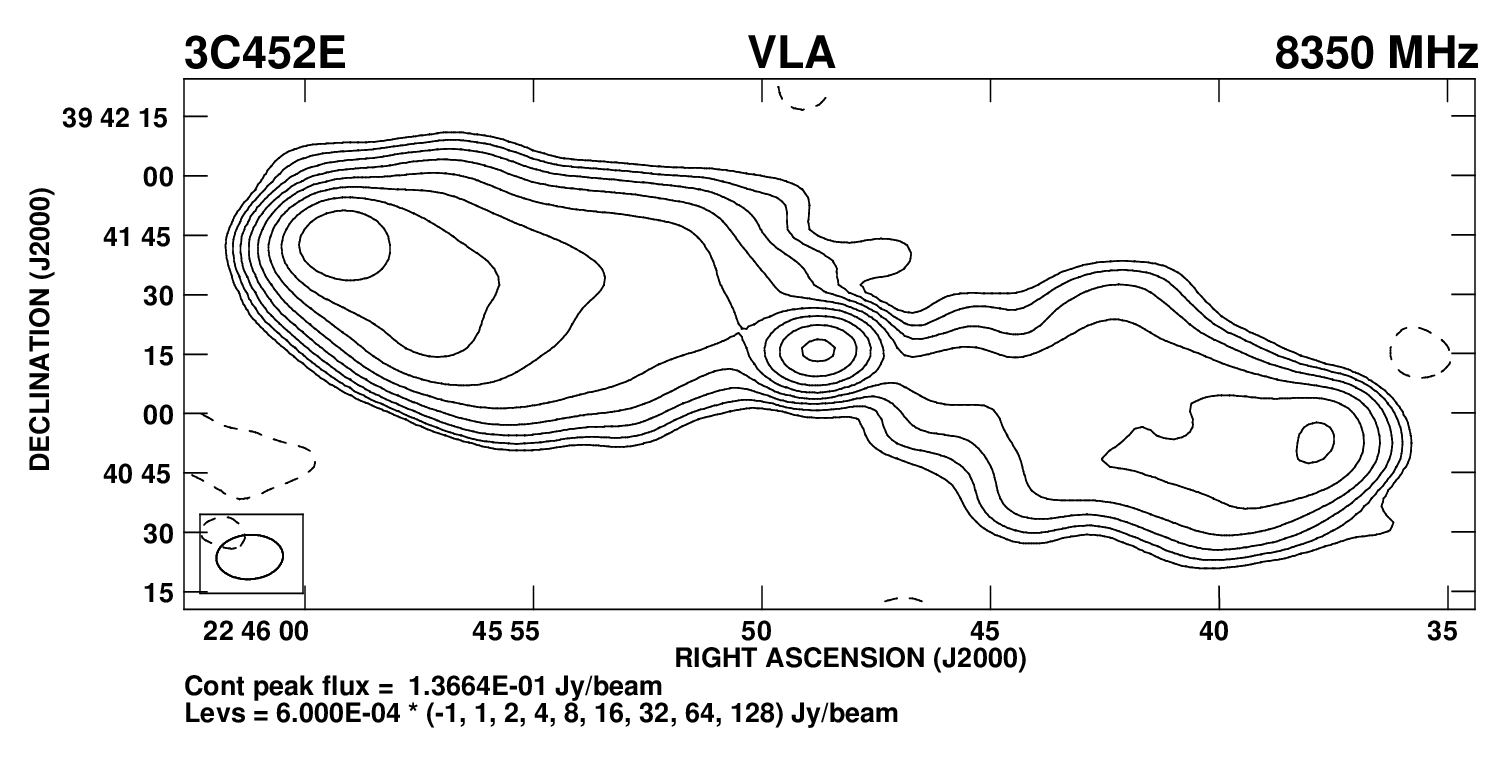,width=3.3in,angle=-90}
     }
\contcaption{}
     \end{figure}

%%%%%%%%%%%%%%%%%%%%%%%%%%%%%%%%%%%%%%%%%%%%%%%%%%%%%%%%%%%%%%%%%%%%%%%%%%%%%%%%%%%%%%%%%%%%%%%%%%%%%%%%%%%%%%%%%%%

%%%%%%%%%%%%%%%%%%%%%%%%%%%%%%%%%%%%%%%%%%%%%%%%%%%%%%%%%%%%%%%%%%%%%%%%%%%%%%%%%%%%%%%%%%%%%%%%%%%%%%%%%%%%%

     \begin{figure}
\vbox{
   \psfig{file=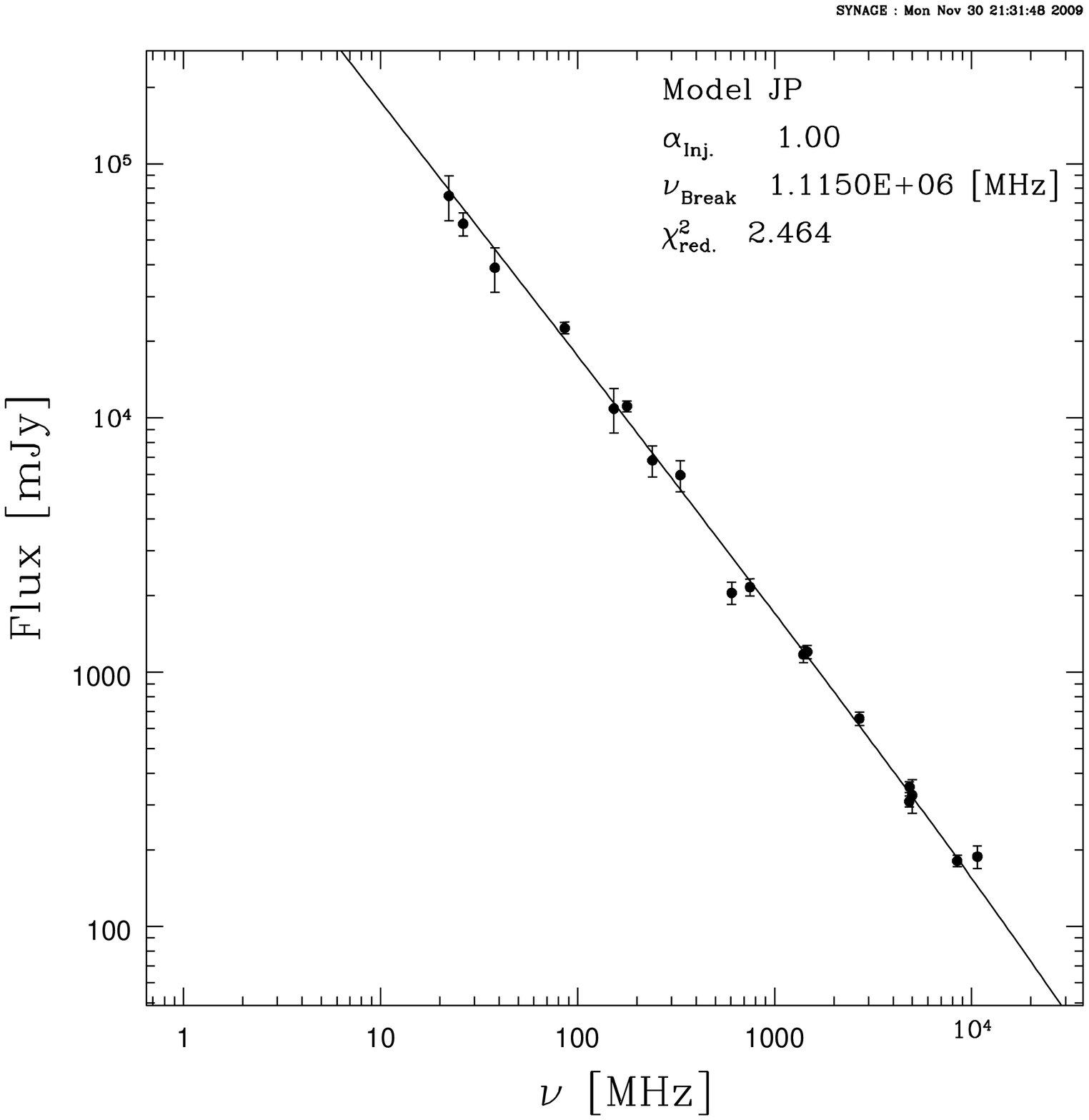,width=3.2in,angle=0}
   \psfig{file=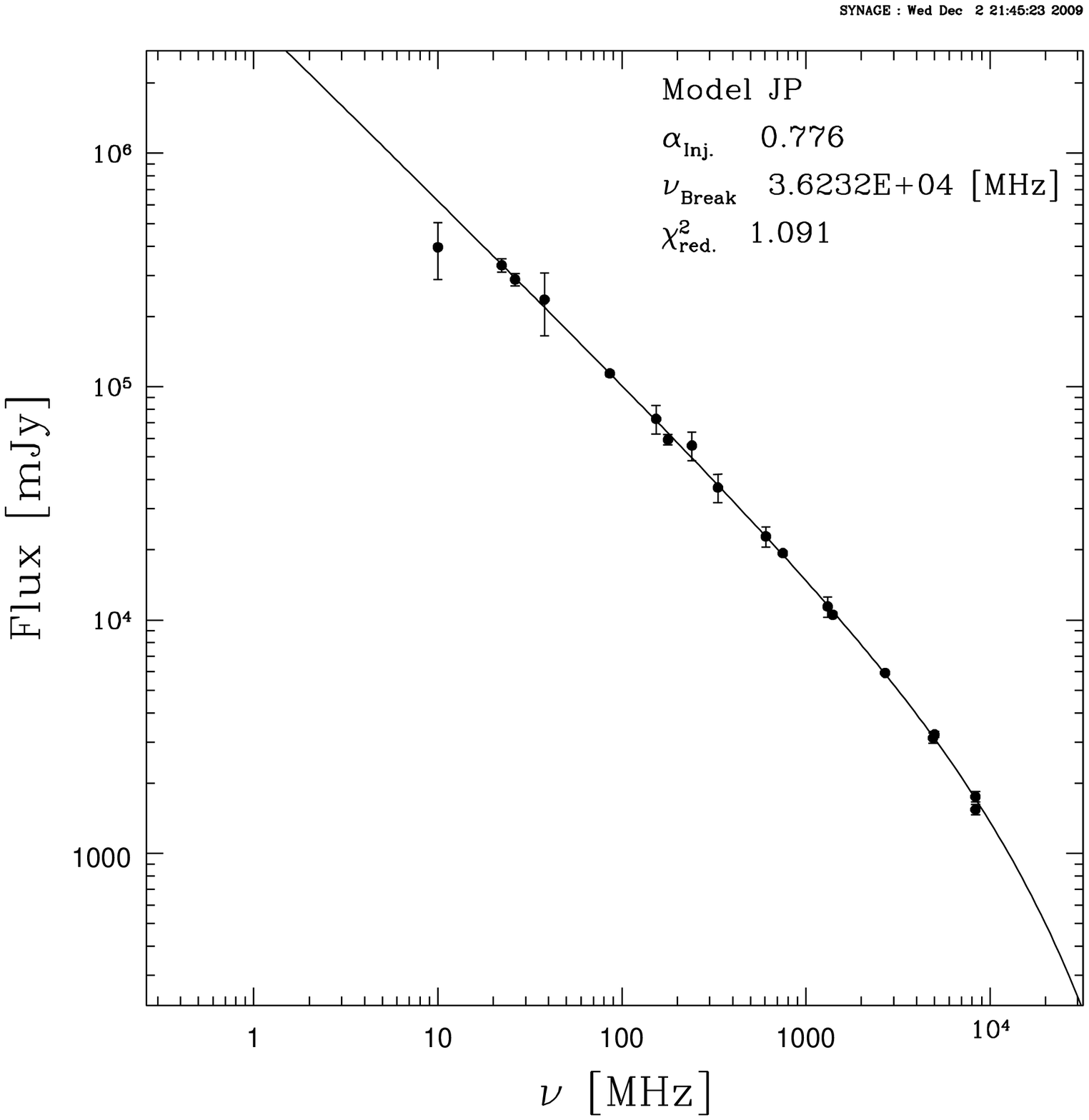,width=3.2in,angle=0}
     }
\caption[]{The spectra of { the extended emission of 3C46 (upper panel) and 3C452 (lower panel) obtained after 
          subtracting the core flux density at frequencies greater than $\sim$1400 MHz from the total flux density.
          Any contributions of the core flux density at lower frequencies are less than $\sim$1 per cent and have been neglected. } 
          The total flux densities are from Laing \& Peacock (1980) and the measurements presented in this paper. 
          { The fits to the spectra obtained using the {\tt SYNAGE} package (Murgia et al. 1999) are also shown.} 
          }
    \end{figure}

%%%%%%%%%%%%%%%%%%%%%%%%%%%%%%%%%%%%%%%%%%%%%%%%%%%%%%%%%%%%%%%%%%%%%%%%%%%%%%%%%%%%%%%%%%%%%%%%%%%%%%%%%%%%%%%%%%%

%%%%%%%%%%%%%%%%%%%%%%%%%%%%%%%%%%%%%%%%%%%%%%%%%%%%%%%%%%%%%%%%%%%%%%%%%%%%%%%
\begin{figure}
\vbox{
   \psfig{file=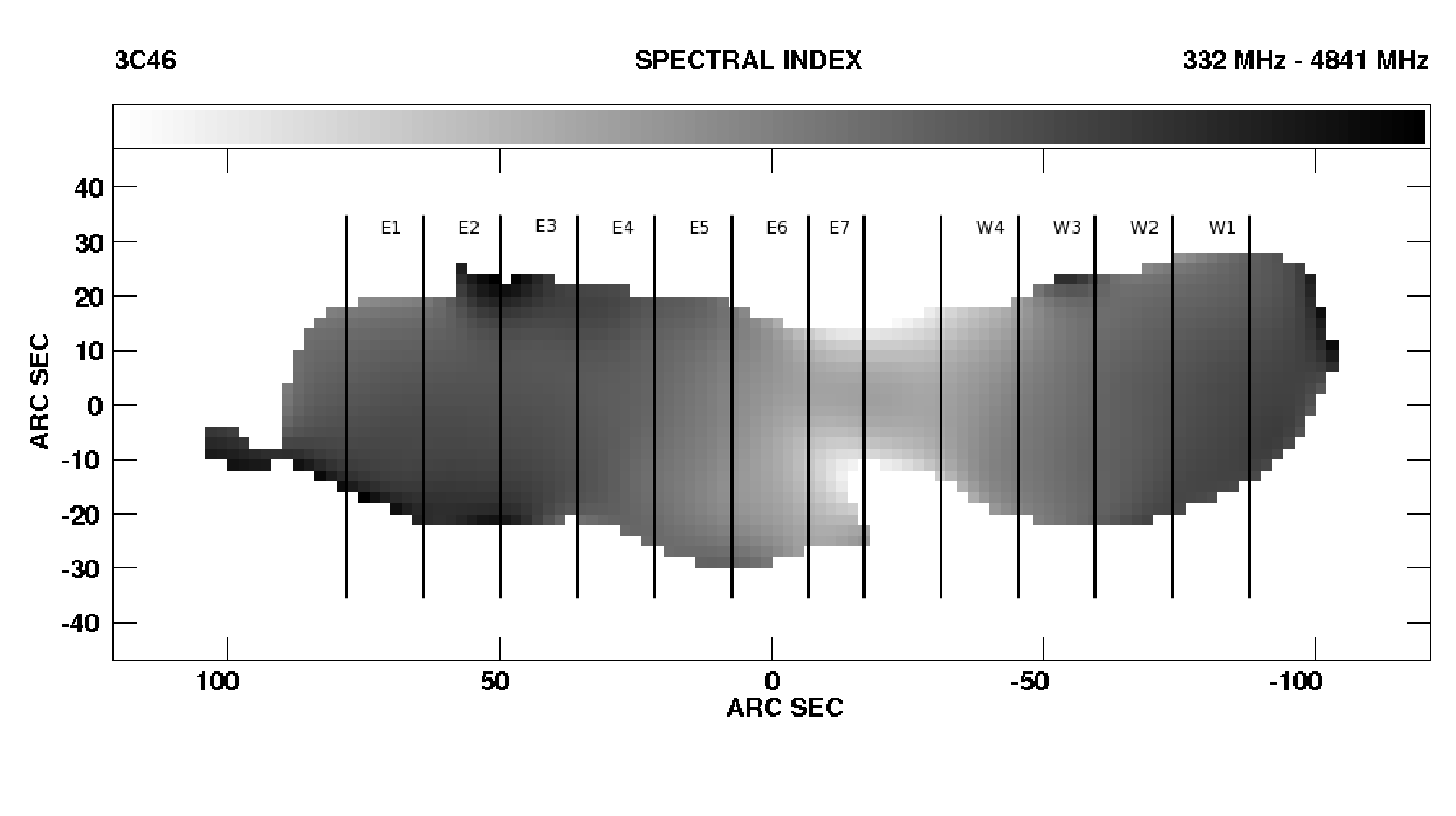,width=3.3in,angle=0}
   \psfig{file=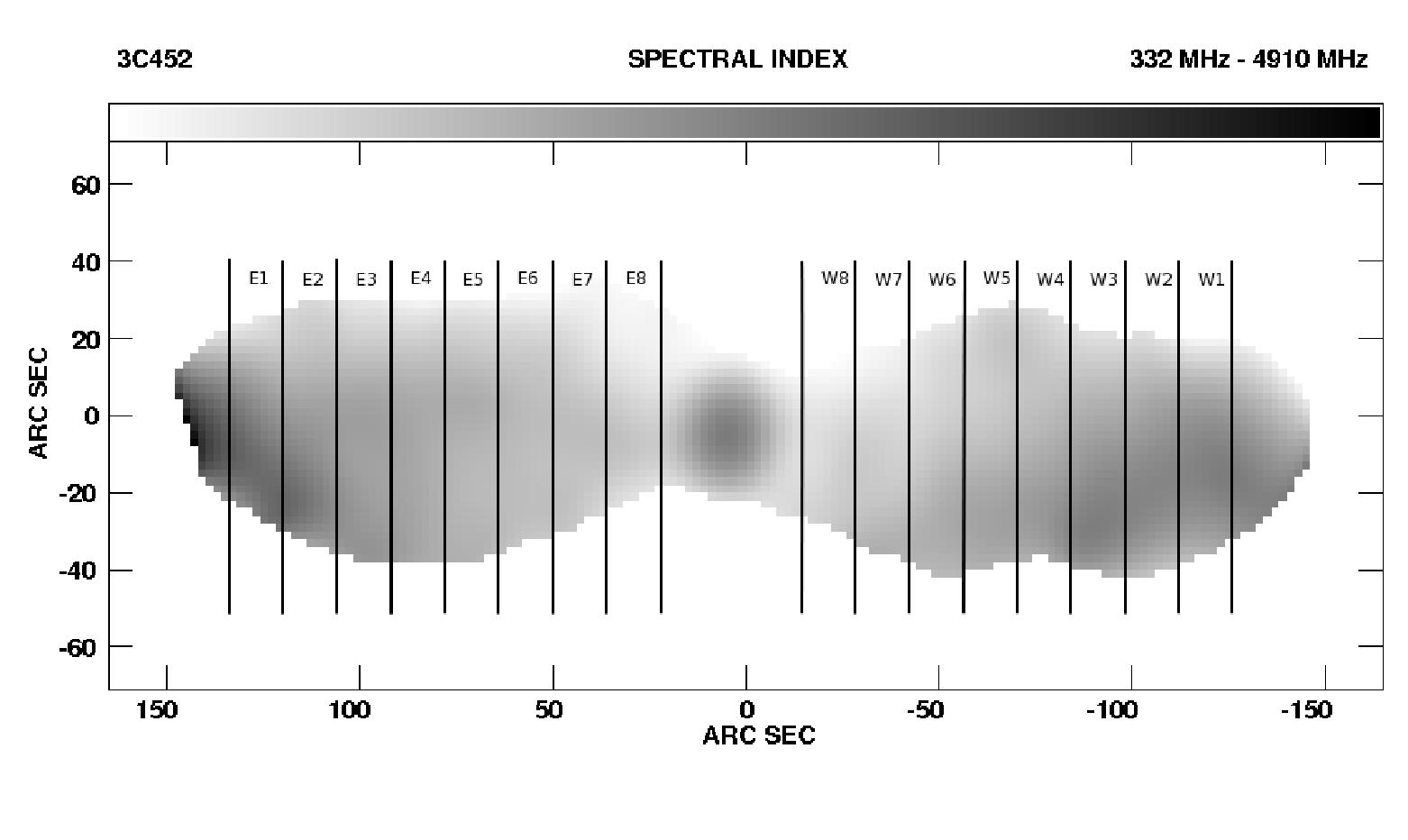,width=3.3in,angle=0}
    }
\caption[]{ { The spectral-index images of 3C46 and 3C452 with the strips used
for estimating the spectral ages along the lobes being marked by vertical lines and
labelled. The central region corresponds to the `Central$-$core' region in Tables 3 and 4
and Fig. 8. The images have been rotated so that they lie in the east-west direction. The 
spectral indices have been estimated between 332 and $\sim$5000 MHz. The grey
scale bar indicates variations in spectral index from 0.4 to 2.0 for 3C46
and $-$0.4 to 1.4 for 3C452. Some of the values at the edges are spurious.}
          }
\end{figure}

%%%%%%%%%%%%%%%%%%%%%%%%%%%%%%%%%%%%%%%%%%%%%%%%%%%%%%%%%%%%%%%%%%%%%%%%%%%%%%% 
\section{Observational results}
The GMRT and VLA images of 3C46 are presented in Fig. 1 while those of 3C452 are presented in 
Fig. 2.  The observational parameters and some of the observed
properties are presented in Table 2, which is arranged as follows.
Column 1: Name of the source; column 2: frequency  of observations in units of MHz, with
the letter G or V representing either GMRT or VLA observations;
columns 3--5: the major and minor axes of the restoring beam in arcsec and its position angle 
(PA) in degrees;
column 6: the rms noise in units of mJy beam$^{-1}$; column 7: the integrated flux density of the
source in mJy. We examined the
change in flux density by specifying different areas around the source 
and found the difference to be within
a few per cent. The flux densities at different frequencies have been estimated over 
similar areas.  Columns 8, 11 and 14: component designation, where W, E and C denote 
the western, eastern and core components respectively;
columns 9 and 10, 12 and 13, and 15 and 16: the peak and total flux densities of each of the
components in units of mJy beam$^{-1}$ and mJy respectively. For the 8350-MHz observations of
3C452, the flux density of the lobe which was at the pointing centre is reliable and has
been listed. The superscript $g$ 
indicates that the flux densities have been estimated from a two-dimensional Gaussian fit to the 
core component.  The { spectra of the extended emission after subtracting the core flux
density at frequencies larger than $\sim$1400 MHz are shown in Fig. 3 along with the fits to
the data using the {\tt SYNAGE} package (Murgia et al. 1999). } The total flux densities are from
Laing \& Peacock (1980) and our measurements.

%%%%%%%%%%%%%%%%%%%%%%%%%%%%%%%%%%%%%%%%%%%%%%%%%%%%%%%%%%%%%%%%%%%%%%%%%%%%%%%%%%%%%%%%%
%%%%%%%%%%%%%%%%%%%%%%%%%%%%%%%%%%%%%%%%%%%%%%%%%%%%%%%%%%%%%%%%%%%%%%%%%%%%%%%%%%%%%%%%%
\section{Discussion and results}

\subsection{Radiative losses}
In the high-luminosity FRII radio sources, as the jets of relativistic plasma
traverse outwards initially through the interstellar medium of the host galaxy
and later through the intracluster and intergalactic medium, they dissipate their
energy at their leading edges. This gives rise to the intense regions of emission
called `hotspots'. The relativistic particles flow out from the hotspots to form the
extended lobes of radio emission, so that the radiating particles closest to the
hotspot are the youngest while those farthest from it are the oldest. The radio continuum
spectra in different parts of an extended radio source contain information about
the various energy losses and gains of the radiating particles during the lifetime
of the source.  If there is no significant
reacceleration within these lobes and no significant mixing of particles, there
should be a spectral gradient across the radio source. The hotspots where the particles are
being accelerated should have the flattest spectral index, reflecting the injection
spectral index $\alpha_{\rm inj}$, while the spectrum should steepen with increasing
distance from the hotspot. Since the high-energy particles lose energy more rapidly,
the steepening in the spectrum would be seen more clearly at high frequencies. This trend has
been reported in several studies and used to estimate the radiative ages and expansion
velocities in the powerful 3CR sources (e.g. Myers \& Spangler 1985;
Alexander \& Leahy 1987; Leahy, Muxlow \& Stephens 1989; Carilli et al. 1991; Liu, Pooley \& Riley 1992),
in the low-luminosity and medium-luminosity radio galaxies (e.g. Klein et al.
1995; Parma et al. 1999), giant radio sources (Konar et al. 2008; Jamrozy et al. 2008 and references
therein) and compact steep-spectrum sources (Murgia et al. 1999). However, there are several
caveats in the interpretation which one needs to bear in mind. These include
details of the backflow of the lobe material, difficulties in disentangling the different
energy losses of the radiating particles and variations of the local magnetic field
(e.g. Wiita \& Gopal-Krishna 1990; Rudnick, Katz-Stone \& Anderson 1994; Eilek
\& Arendt 1996; Jones, Ryu \& Engel 1999; Blundell \& Rawlings 2000).

%%%%%%%%%%%%%%%%%%%%%%%%%%%%%%%%%%%%%%%%%%%%%%%%%%%%%%%%%%%%%%%%%%%%%%%%%%%%%%%%%%%%%%%%%%%%%%%%%%%%%%%%%%%%%%%%%%%
\begin{table*}
\caption{The observational parameters and observed properties of the sources}

\begin{tabular}{l l rrr r r r rr l rr r rr}
\hline
Source   & Freq.       & \multicolumn{3}{c}{Beam size}                    & rms      & S$_I$   & Cp  & S$_p$  & S$_t$  & Cp   & S$_p$ & S$_t$ & Cp  & S$_p$   & S$_t$     \\

           & MHz         & $^{\prime\prime}$ & $^{\prime\prime}$ & $^\circ$ &    mJy   & mJy     &     & mJy    & mJy    &      & mJy   & mJy   &     & mJy     & mJy       \\
           &             &                   &                   &          & beam$^{-1}$&       &     &beam$^{-1}$   &        &      & beam$^{-1}$    &       &     & beam$^{-1}$      &           \\ 
   (1)     & (2)   & (3)  & (4)  & (5)  & (6)  & (7)  &(8)& (9)  & (10) & (11)  &   (12)  &(13)  &(14)& (15) & (16)  \\
\hline
3C46      & G153  & {    28.5}  & {    16.0} & {    65}   & {    6.9}   & {    10880}    & W  & {    2325}    & {  6029}  & C    &       &     &  E  &  {    1852}  & {    4899}  \\
          & G240  & 13.2  &  8.1 & 48   & 4.3   & 6806    & W  &  771    &  3821  & C    &       &     &  E  &   721  & 3099  \\
          & G332  & 10.3  &  9.1 & 69   & 0.54  & 5957    & W  &  725    &  3172  & C    &       &     &  E  &   734  & 2689  \\
          & G606  &  5.2  &  3.8 & 48   & 0.23  & 2051    & W  &  157    &  1537  & C    &       &     &  E  &   277  & 1402  \\
          & V1465 & 13.8  & 12.7 & 12   & 0.61  & 1202    & W  &  253    &   640  & C    &       &     &  E  &   197  &  590  \\
          & V4841 & 13.9  & 11.6 & 166  & 0.15  &  355    & W  &   76    &   185  & C$^g$&  1.2  & 1.1 &  E  &    60  &  172  \\
          & V8460 &  7.7  &  6.9 & 32   & 0.04  &  183    & W  &   25    &    88  & C$^g$&  1.6  & 1.8 &  E  &    30  &   94  \\

3C452     & G153   &{    24.1} &{    18.1} & {    32}  & {    8.7}  &{    81142}   & W  & {    4341}   &{    38725}  & C    &       &     &  E  & {    3909}  &{    42366} \\
          & G240   & 13.1 & 11.8 &  34  &  4.5  & 55706   & W &{    1810}& 27589  & C    &       &     &  E&{    1825}& 28422  \\
          & G332   & 14.9 &  8.3 & 134  &  2.6  & 37703   & W  &   967   & 18191  & C    &       &     &  E  &   862  & 18724  \\
          & G606   &  6.7 &  4.4 &  98  &  0.68 & 24369   & W  &   207   & 11246  & C    &       &     &  E  &   251  & 11893  \\
          & G1313  & 16.4 & 10.5 &  65  &  0.92 & 11565   & W  &   518   &  5110  & C$^g$ & 213  & 203 &  E  &   503  &  5677  \\
          & V4910  & 17.0 & 12.6 &  85  &  0.44 &  3318   & W  &   209   &  1435  & C$^g$ & 153  & 213 &  E  &   188  &  1399  \\
          & V8350W & 16.8 & 11.7 &  94  &  0.06 &{    1860}& W &   151   &   860  & C$^g$ & 121  & 145 &  E  &        &        \\
          & V8350E & 16.8 & 11.2 &  93  &  0.07 &  1865   & W  &         &        & C$^g$ & 134  & 151 &  E  &   137  &   816  \\
\hline       
\end{tabular}
\end{table*}
%%%%%%%%%%%%%%%%%%%%%%%%%%%%%%%%%%%%%%%%%%%%%%%%%%%%%%%%%%%%%%%%%%%%%%%%%%%%%%%%%%%%%%%%%%%%%%%%%%%%%%%%%%%%%%%%%%%

%%%%%%%%%%%%%%%%%%%%%%%%%%%%%%%%%%%%%%%%%%%%%%%%%%%%%%%%%%%%%%%%%%%%%%%%%%%%%%%%%%%%%%%%%%%%%%%%%%%%%%%%%%%%%%%%
\begin{figure*}
\vbox{
   \hbox{
      \psfig{file=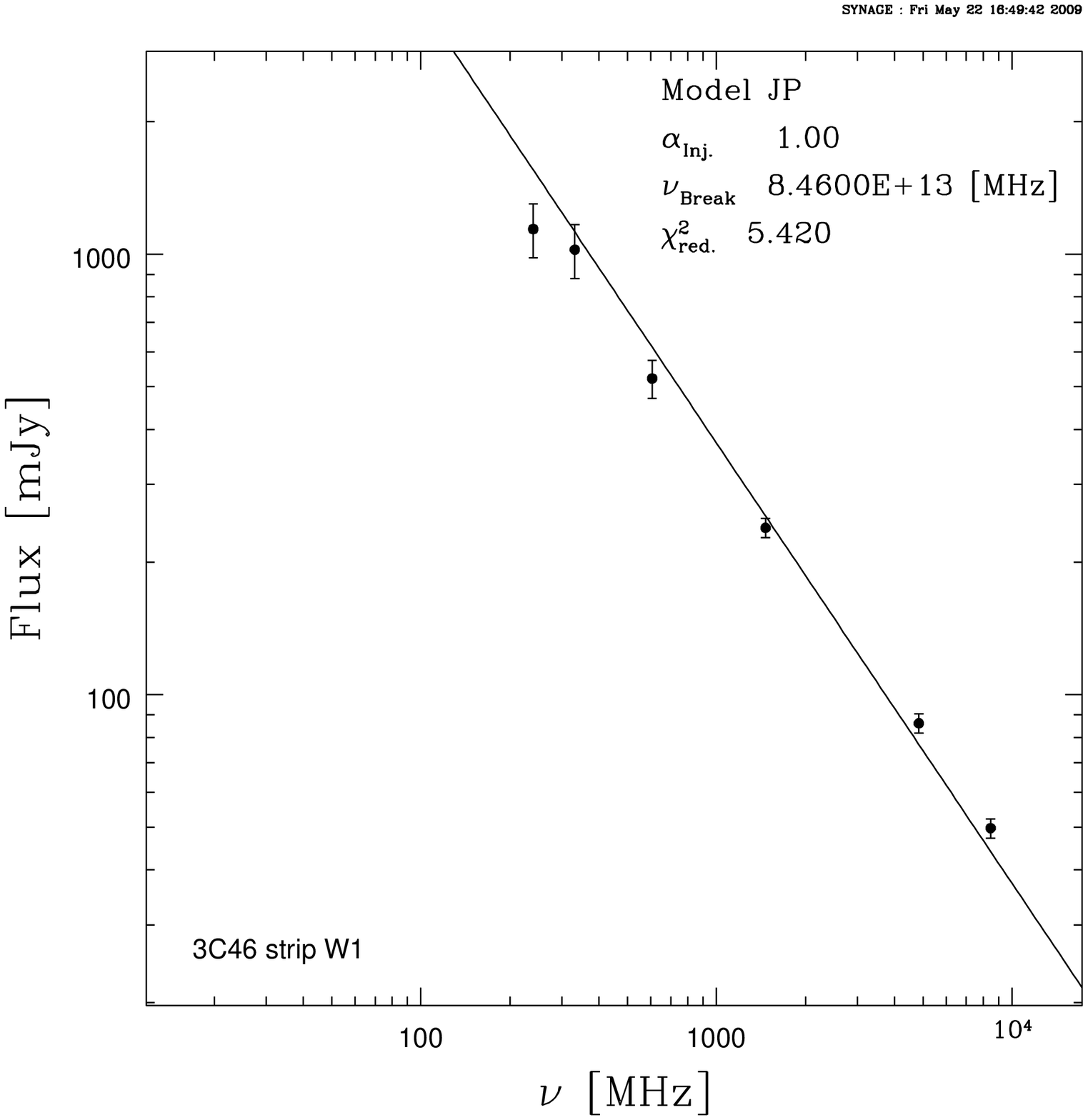,width=2.3in,angle=0}
      \psfig{file=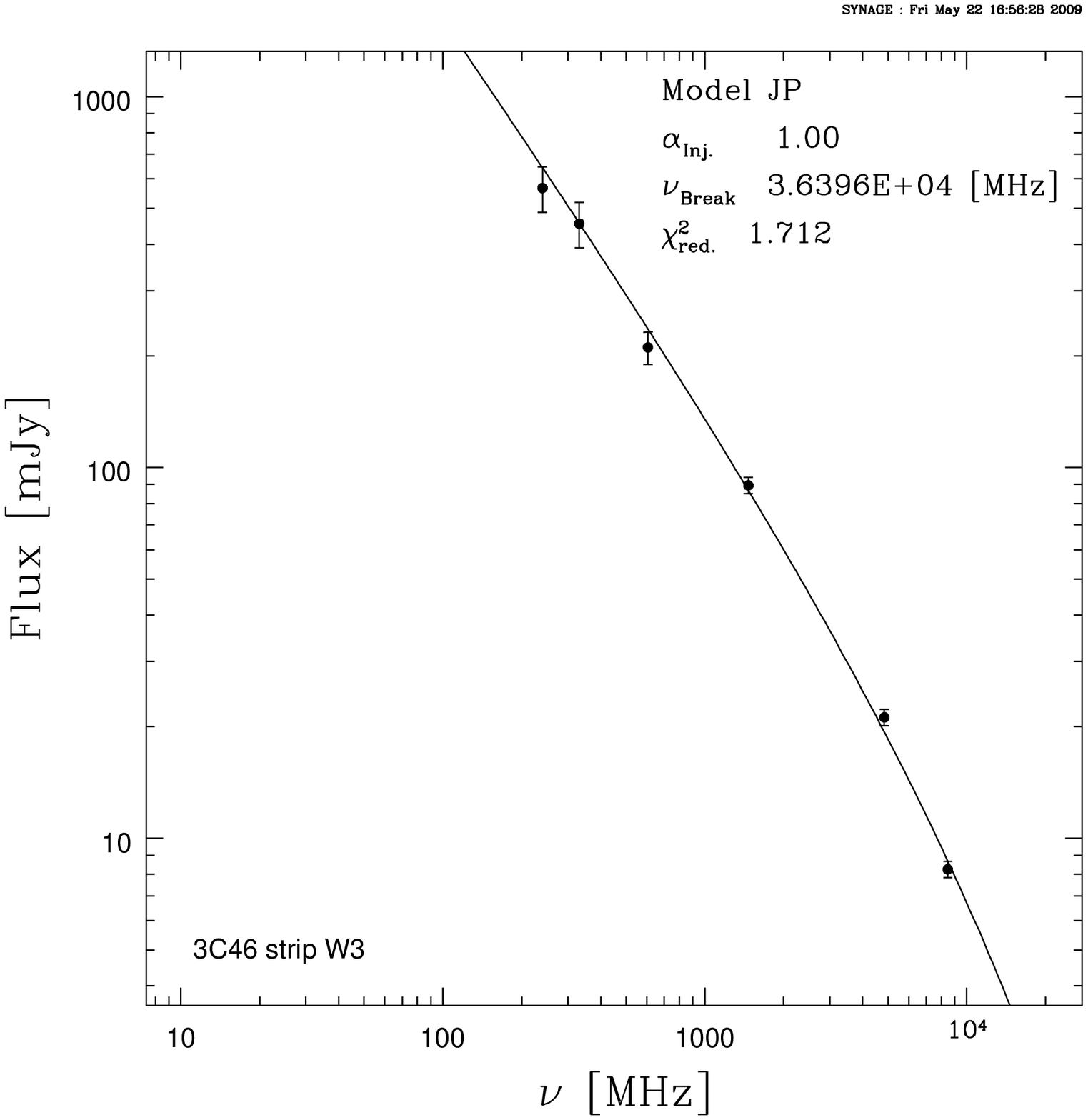,width=2.3in,angle=0}
      \psfig{file=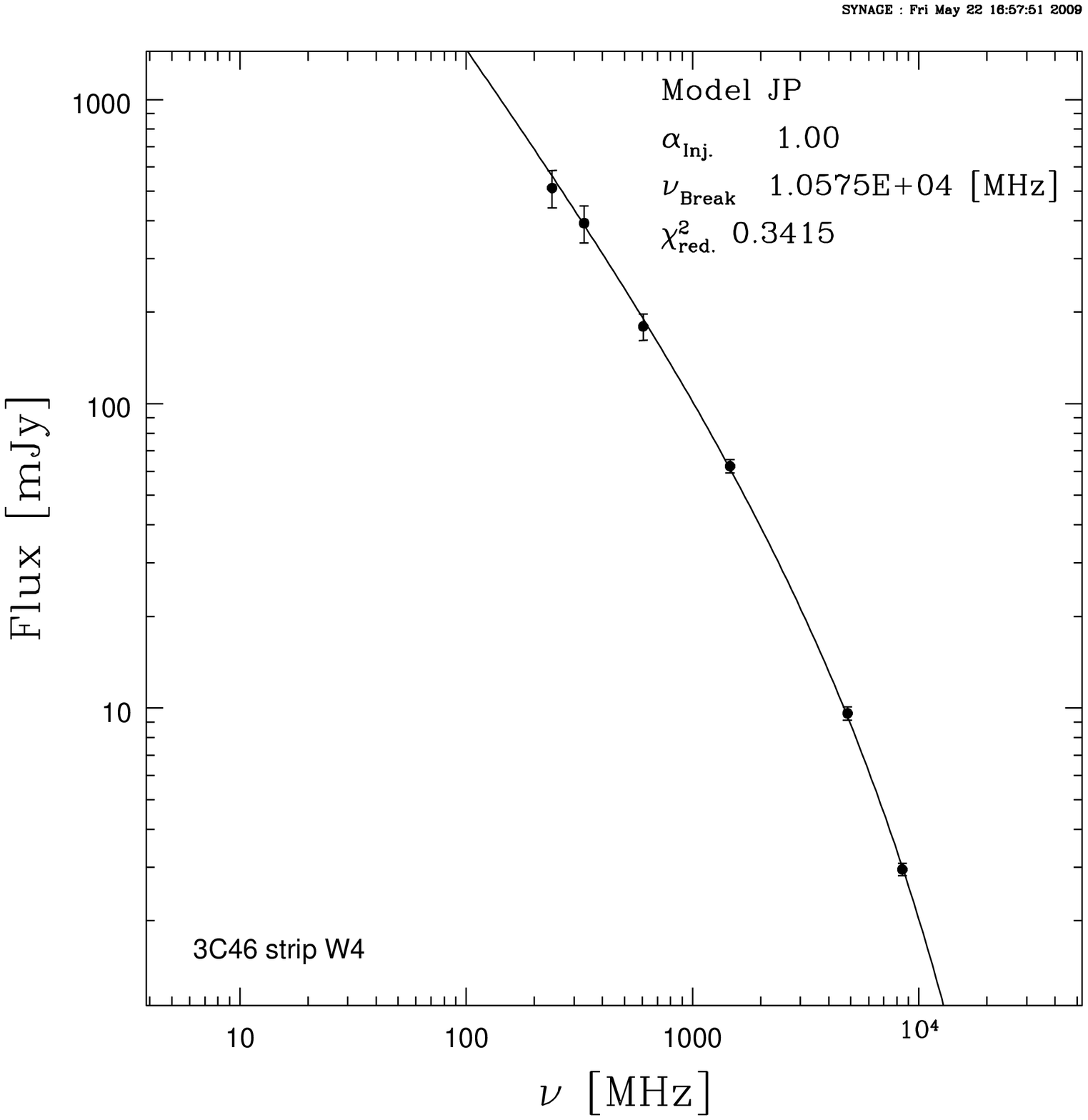,width=2.3in,angle=0}
        }
   \hbox{
      \psfig{file=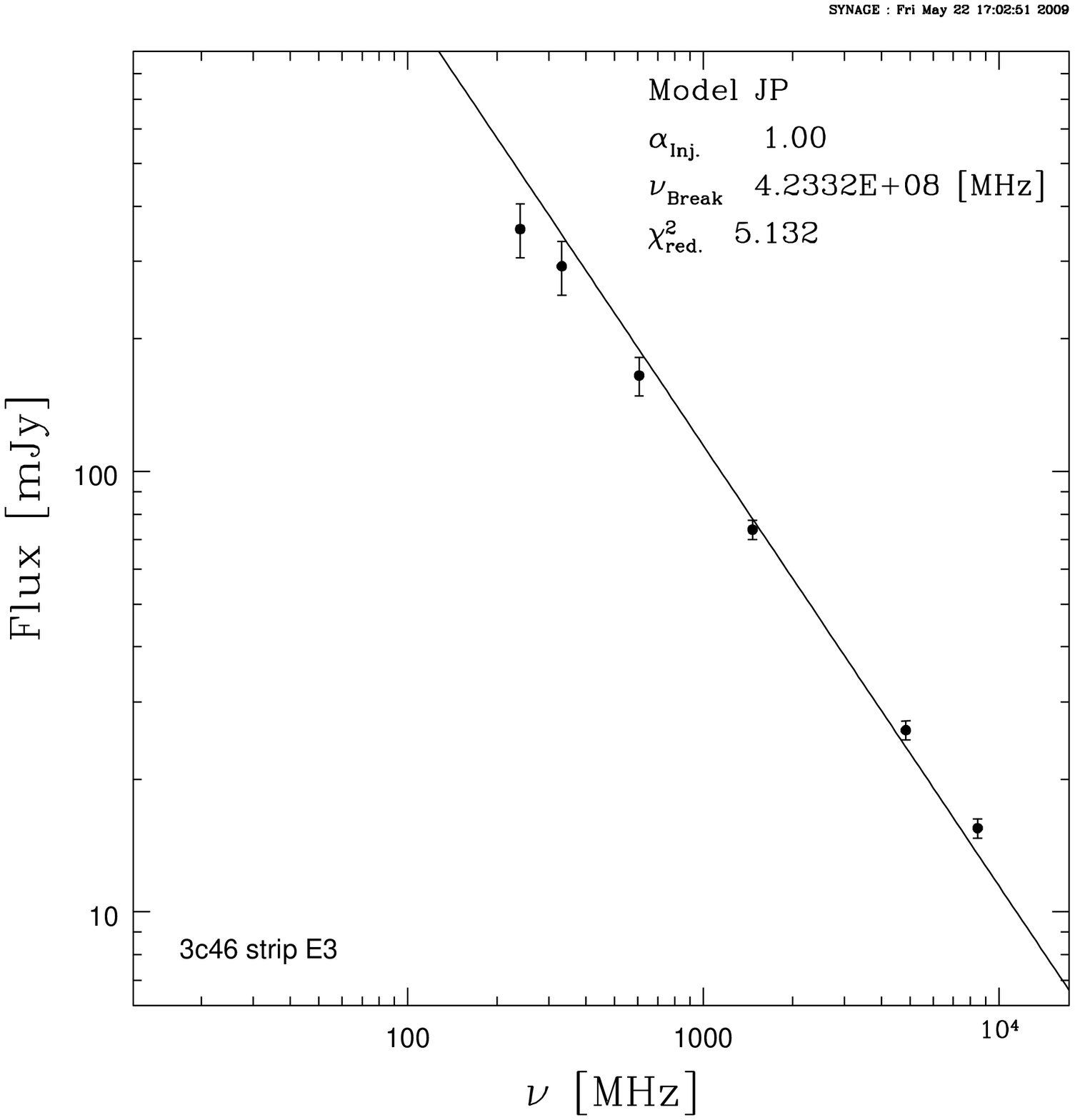,width=2.3in,angle=0}
      \psfig{file=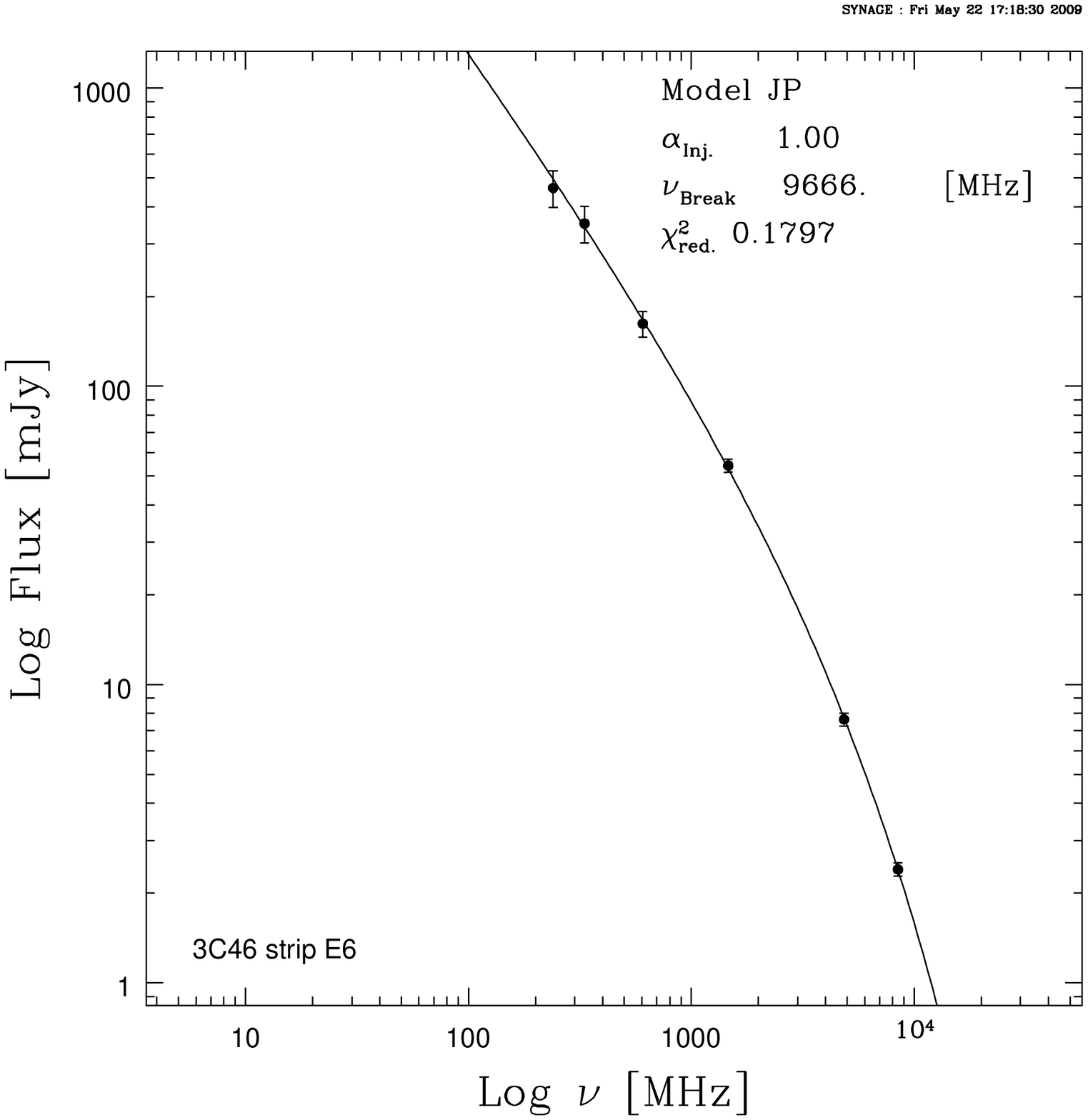,width=2.3in,angle=0}
      \psfig{file=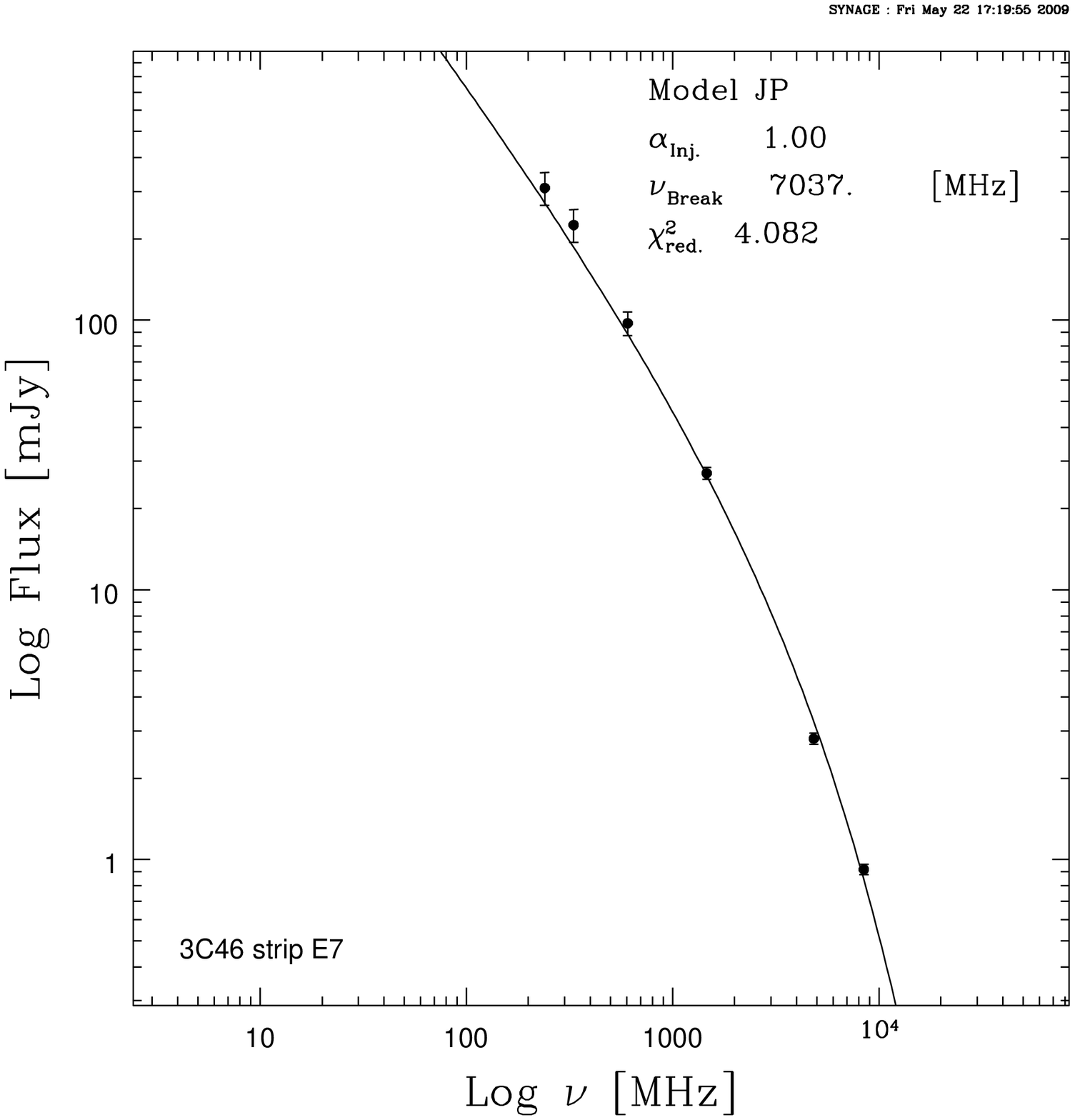,width=2.3in,angle=0}
        }
     }
\caption[]{Typical spectra of the strips for the western (upper panel) and eastern (lower panel) lobes of 3C46,
           with the fits from the JP model as described in the text.}
\end{figure*}
%%%%%%%%%%%%%%%%%%%%%%%%%%%%%%%%%%%%%%%%%%%%%%%%%%%%%%%%%%%%%%%%%%%%%%%%%%%%%%%%%%%%%%%%%%%%%%%%%%%%%%%%%%%%%%%%

%%%%%%%%%%%%%%%%%%%%%%%%%%%%%%%%%%%%%%%%%%%%%%%%%%%%%%%%%%%%%%%%%%%%%%%%%%%%%%%%%%%%%%%%%%%%%%%%%%%%%%%%%%%%%%%%
\begin{figure*}
\vbox{
   \hbox{
      \psfig{file=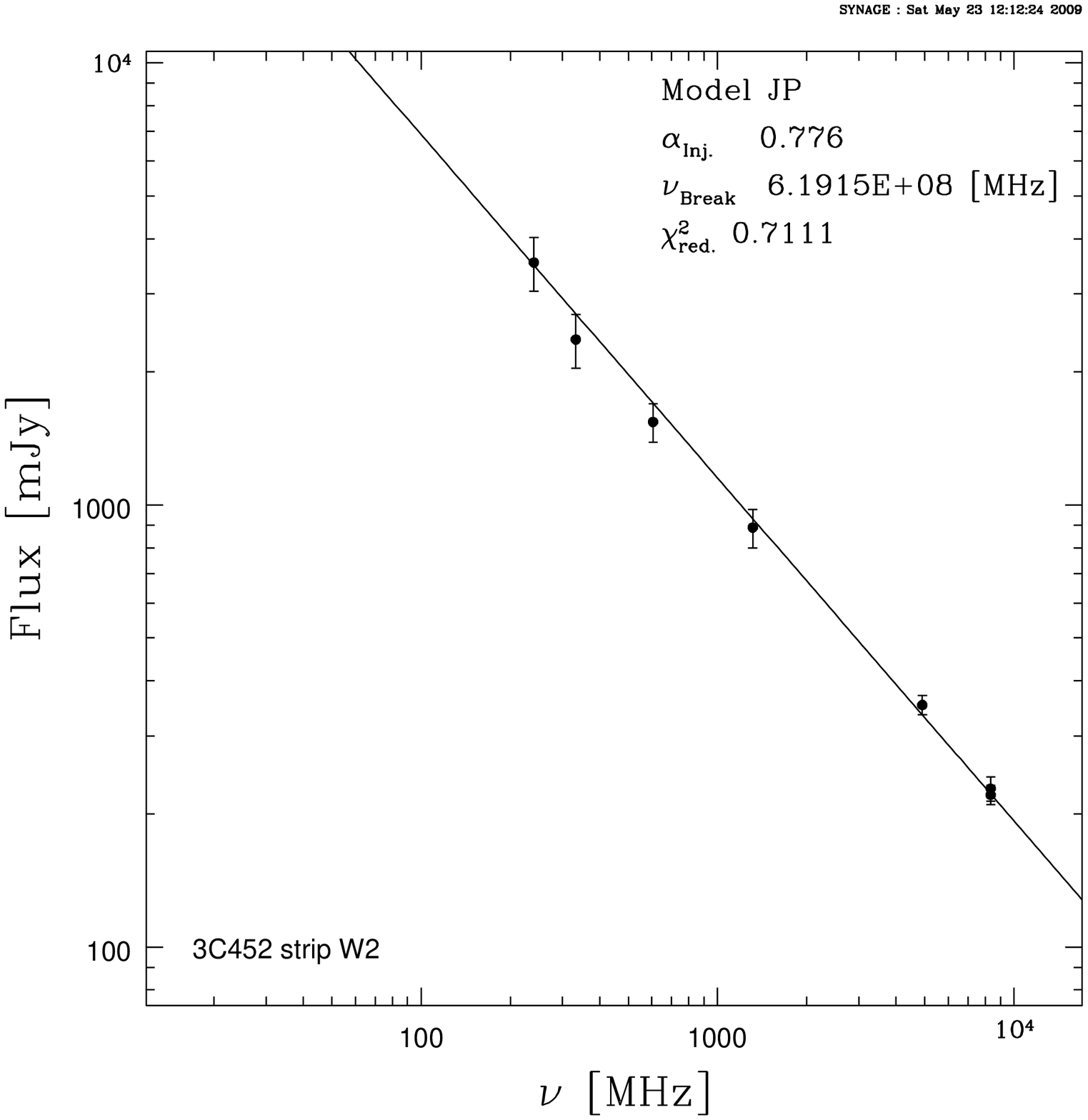,width=2.3in,angle=0}
      \psfig{file=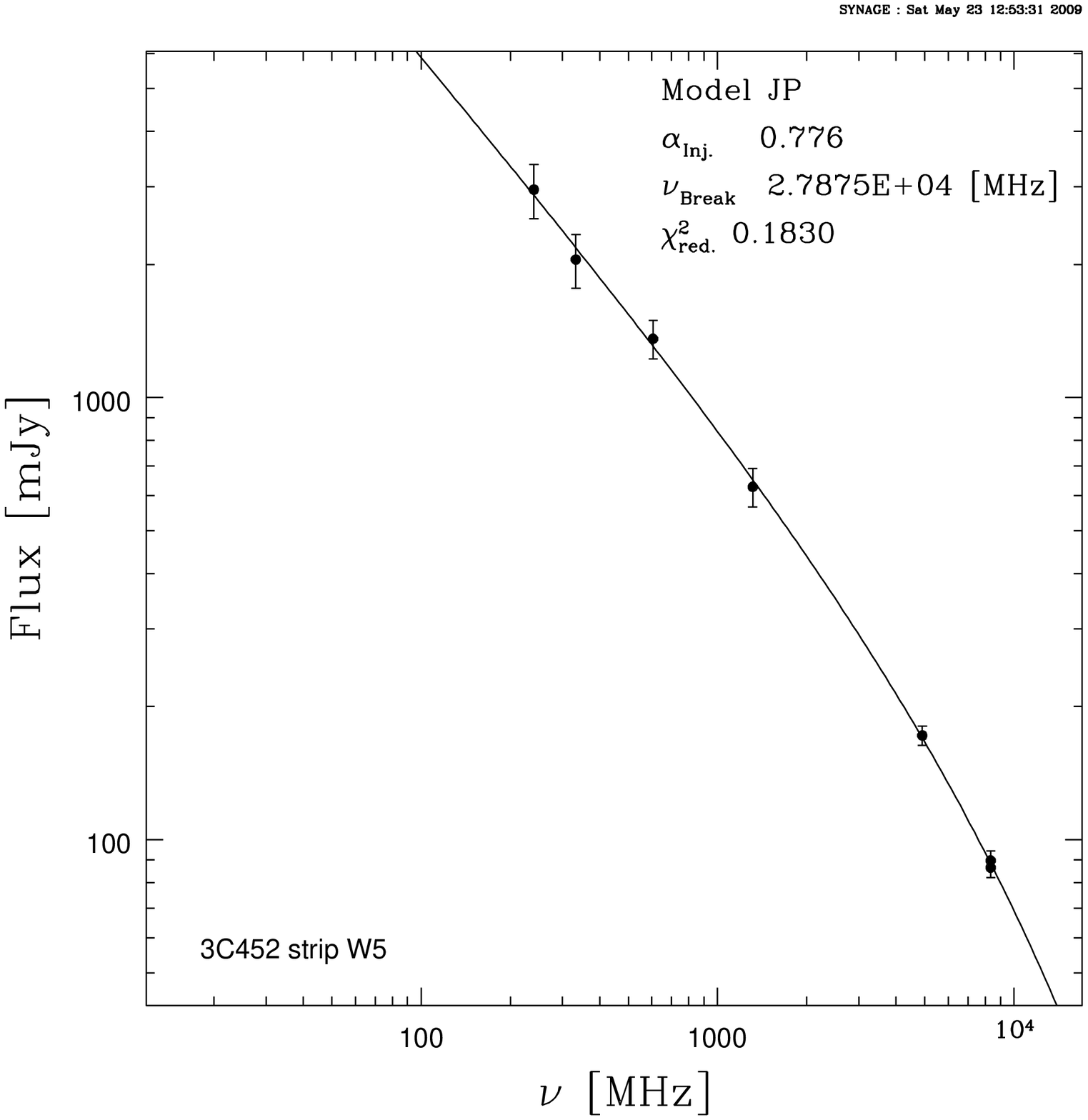,width=2.3in,angle=0}
      \psfig{file=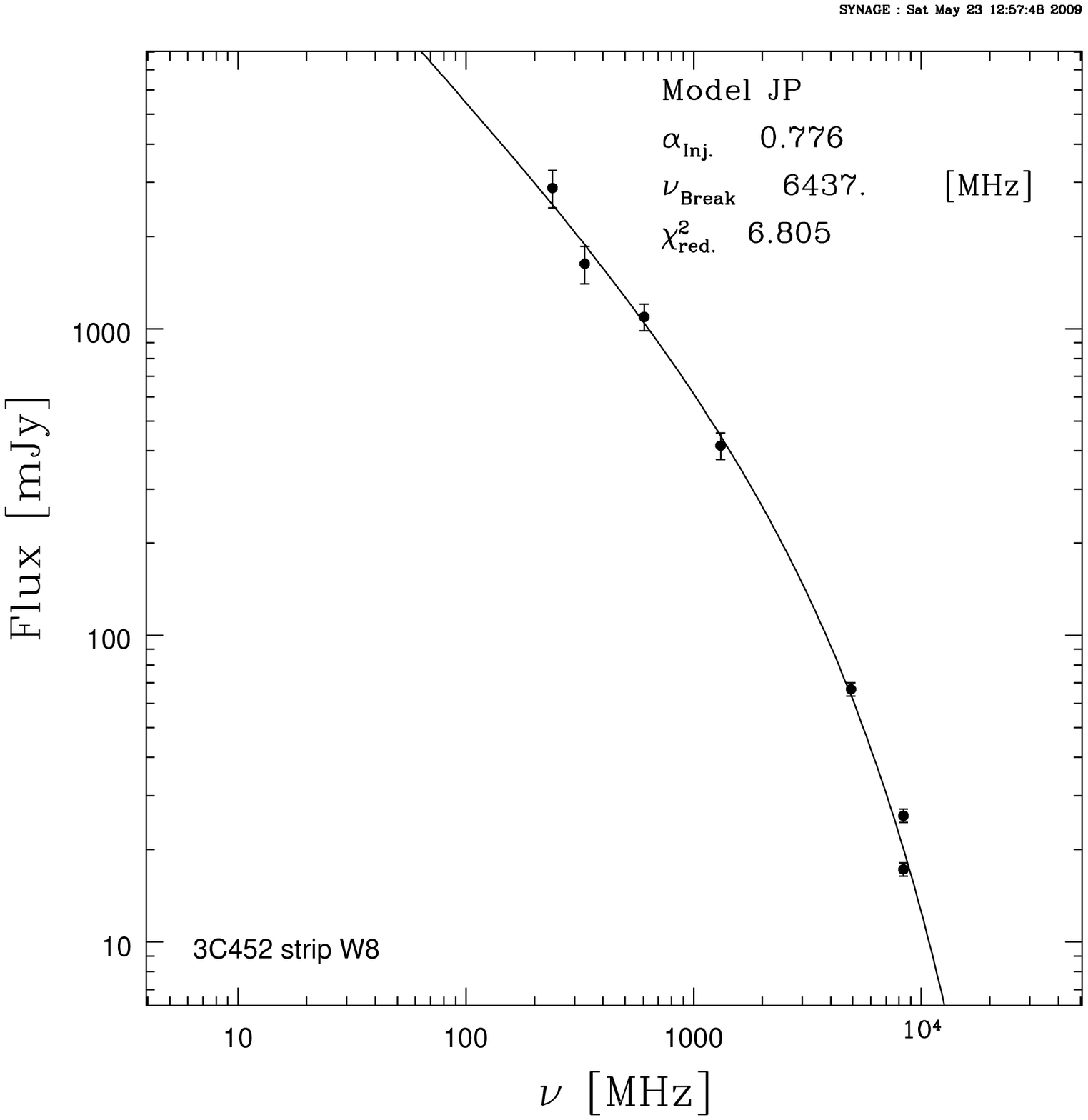,width=2.3in,angle=0}
        }
   \hbox{
      \psfig{file=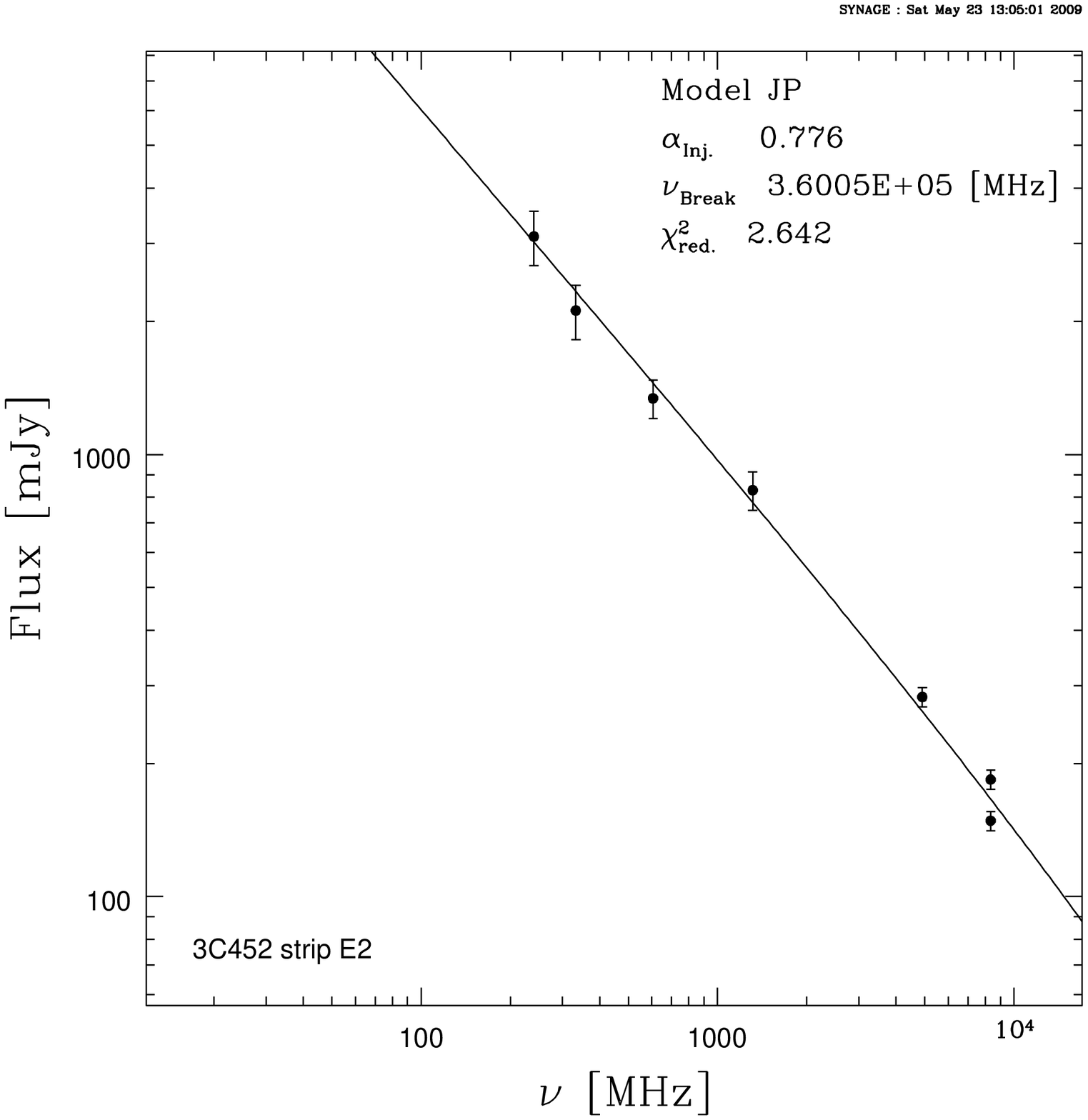,width=2.3in,angle=0}
      \psfig{file=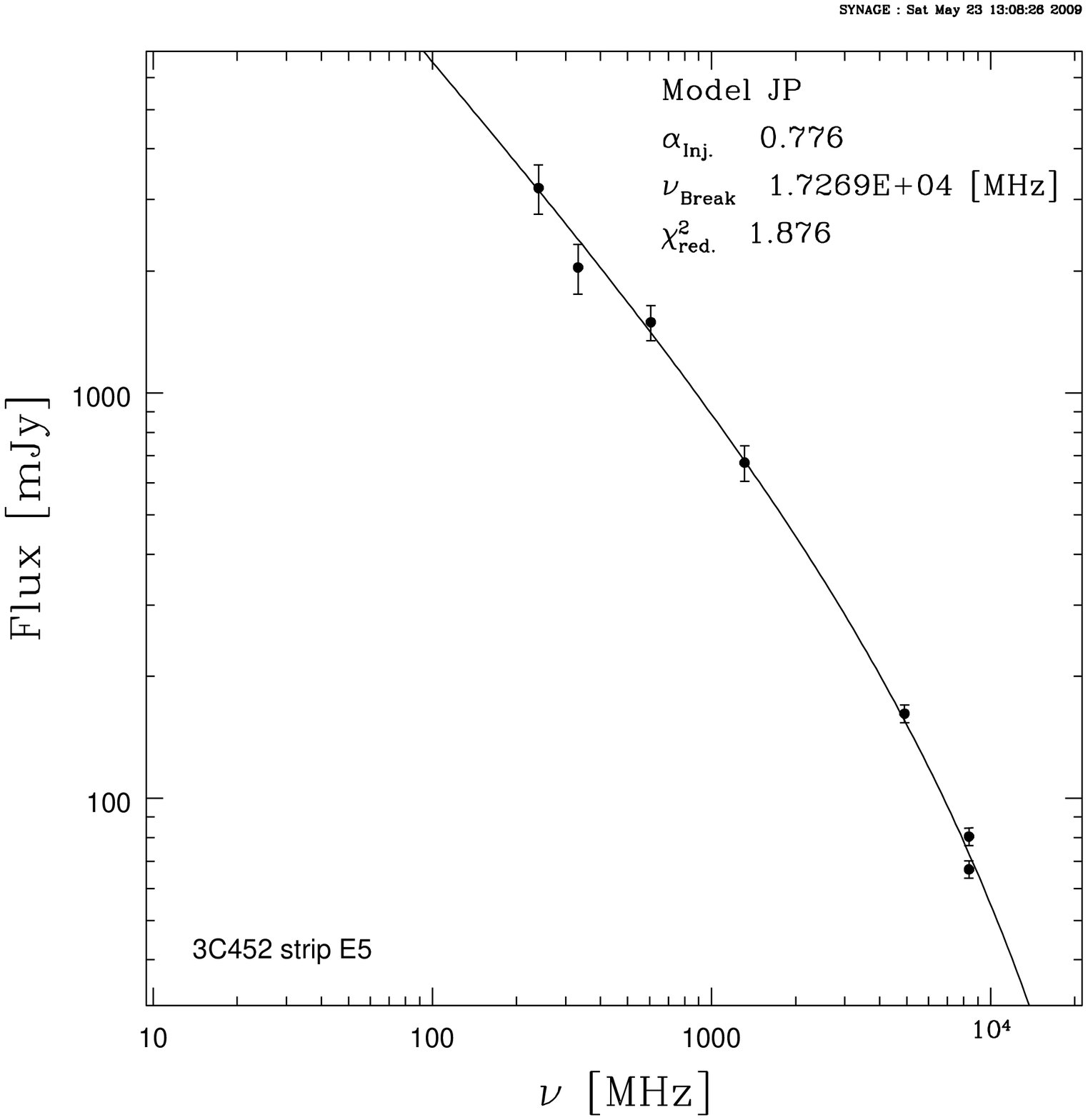,width=2.3in,angle=0}
      \psfig{file=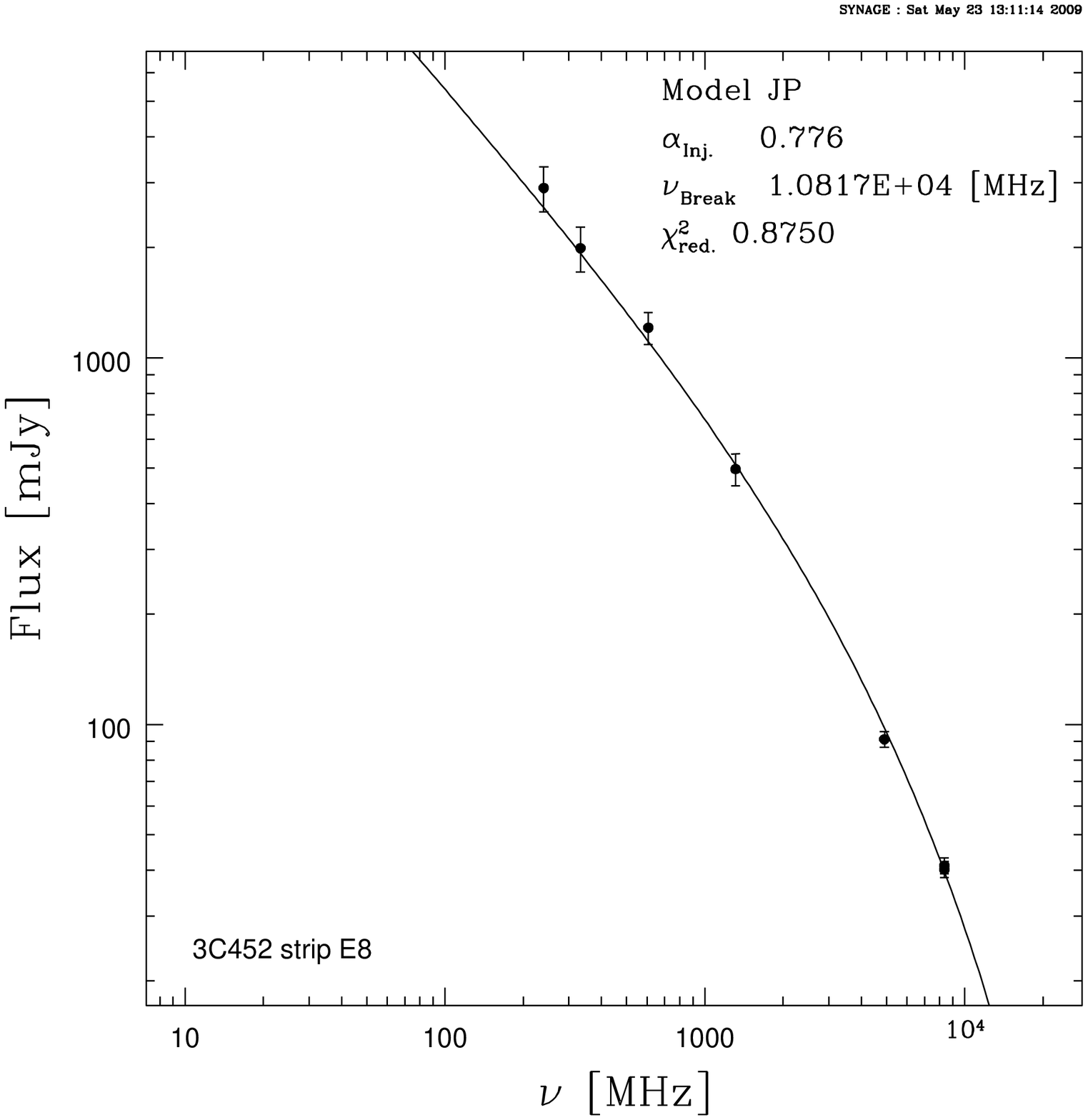,width=2.3in,angle=0}
        }
     }
\caption[]{Typical spectra of the strips for the western (upper panel) and eastern (lower panel) lobes of 3C452,
           with the fits from the JP model as described in the text. }
\end{figure*}
%%%%%%%%%%%%%%%%%%%%%%%%%%%%%%%%%%%%%%%%%%%%%%%%%%%%%%%%%%%%%%%%%%%%%%%%%%%%%%%%%%%%%%%%%%%%%%%%%%%%%%%%%%%%%%%%

Nevertheless, these provide useful inputs towards understanding the different physical
processes which play a role in the evolution of these radio sources. The large radio sources are
particularly suitable for the classical spectral-ageing analysis due to their large angular extent
which can be covered by a significant number of resolution elements. Combining low-frequency
information from the GMRT along with high-frequency ones from the VLA is likely to yield
the most reliable estimates of the spectral break frequency, as has been demonstrated in
a number of recent studies (Konar et al. 2008; Jamrozy et al. 2008 and references therein).

%%%%%%%%%%%%%%%%%%%%%%%%%%%%%%%%%%%%%%%%%%%%%%%%%%%%%%%%%%%%%%%%%%%%%%%%%
\subsection{Spectral ageing analysis}

The observed spectra have been fitted using the
Jaffe \& Perola (1973, JP) and the Kardashev-Pacholczyk (Kardashev 1962;
Pacholczyk 1970; KP) models using the {\tt SYNAGE} package (Murgia et al. 1999).
As reported by Jamrozy et al. (2008), there is no significant difference 
between these two models over the frequency range of our observations, and the
JP model tends to give a better fit to the different strips in the lobes than with
the continuous injection (Kardashev 1962; CI) model. The CI model sometimes
gives a somewhat better fit in the area of a prominent hotspot, but since with
the resolution of our observations the flux density of the hotspots are 
contaminated by lobe emission it does not make a significant difference.

Assuming that (i) the magnetic field strength in a given lobe is
constant throughout the energy-loss process, (ii) the particles injected into the
lobe have a constant power-law energy spectrum with an index $\gamma$, and
(iii) the time-scale of isotropization of the pitch angles of the particles is short
compared with their radiative lifetime, the spectral age, $\tau_{\rm spec}$, is given by

\begin{equation}
\tau_{\rm spec}=50.3\frac{B^{1/2}}{B^{2}+B^{2}_{\rm iC}}\left\{\nu_{\rm br}(1+z)\right\}
^{-1/2} [{\rm Myr}],
\end{equation}

\noindent
where $B_{\rm iC}$=0.318(1+$z$)$^{2}$ is the magnetic field strength equivalent to
the cosmic microwave background radiation. Here $B$, the magnetic field strength of the
lobes, and $B_{\rm iC}$ are expressed
in units of nT, $\nu_{\rm br}$ is the spectral break frequency in GHz above which the
radio spectrum steepens from the initial power-law spectrum given by
$\alpha_{\rm inj}$=$(\gamma-1)/2$. Alexander \& Leahy (1987) and
Alexander (1987) have suggested that the effects of expansion losses may be neglected.

%%%%%%%%%%%%%%%%%%%%%%%%%
To estimate the values of $\alpha_{\rm inj}$ for the two sources, we fit the JP model
to the flux densities of the entire lobes, treating $\alpha_{\rm inj}$ as a free parameter
as well as the total flux density measurements of the source which go to lower frequencies 
(Laing \& Peacock 1980) than our measurements. These yield injection spectral indices of
1.00 and 0.78 for 3C46 and 3C452 respectively (Fig. 3).
Having estimated the $\alpha_{\rm inj}$ values, the total-intensity images of 3C46 and
3C452 have been convolved to a common resolution of { 14 and 17 arcsec respectively
to be consistent with the lowest-resolution image
for all images at $\sim$240 MHz and above.} The 150-MHz images have not been used for this
analysis since their resolution is coarser by another factor of two, and we wish to 
have at least $\sim${ 10} resolution elements along the axis of the source.
Each lobe is then split into a number of strips { as shown in Fig. 4,
separated approximately by the resolution element along the axis of the
source, and also ensuring that the core component lies between two vertical lines and
can be subtracted reliably.} The extreme strips are centred at the peaks of brightness
on the convolved maps. Both these sources have prominent bridges of emission, and we
have also fitted the spectrum to the central region of the source after subtracting
the flux density of the radio core.  Using the {\tt SYNAGE} software we determine the best
fit to the spectrum in each strip from $\sim$240 to 8000 MHz using the JP model, and 
derive the relevant value of $\nu_{\rm br}$. A few examples of the fits to the 
different strips of both the lobes in 3C46 and 3C452 are presented in Figs. 5 and 6
respectively, while the fits to the central regions of the source, which as expected
show the lowest values of $\nu_{\rm br}$ are presented in Fig. 7. 

%%%%%%%%%%%%%%%%%%%%%%%%%%%%%%%%%%%%%%%%%%%%%%%%%%%%%%%%%%%%%%%%%%%%%%%%%%%%%%%%%%%%
\begin{figure}
\vbox{\psfig{file=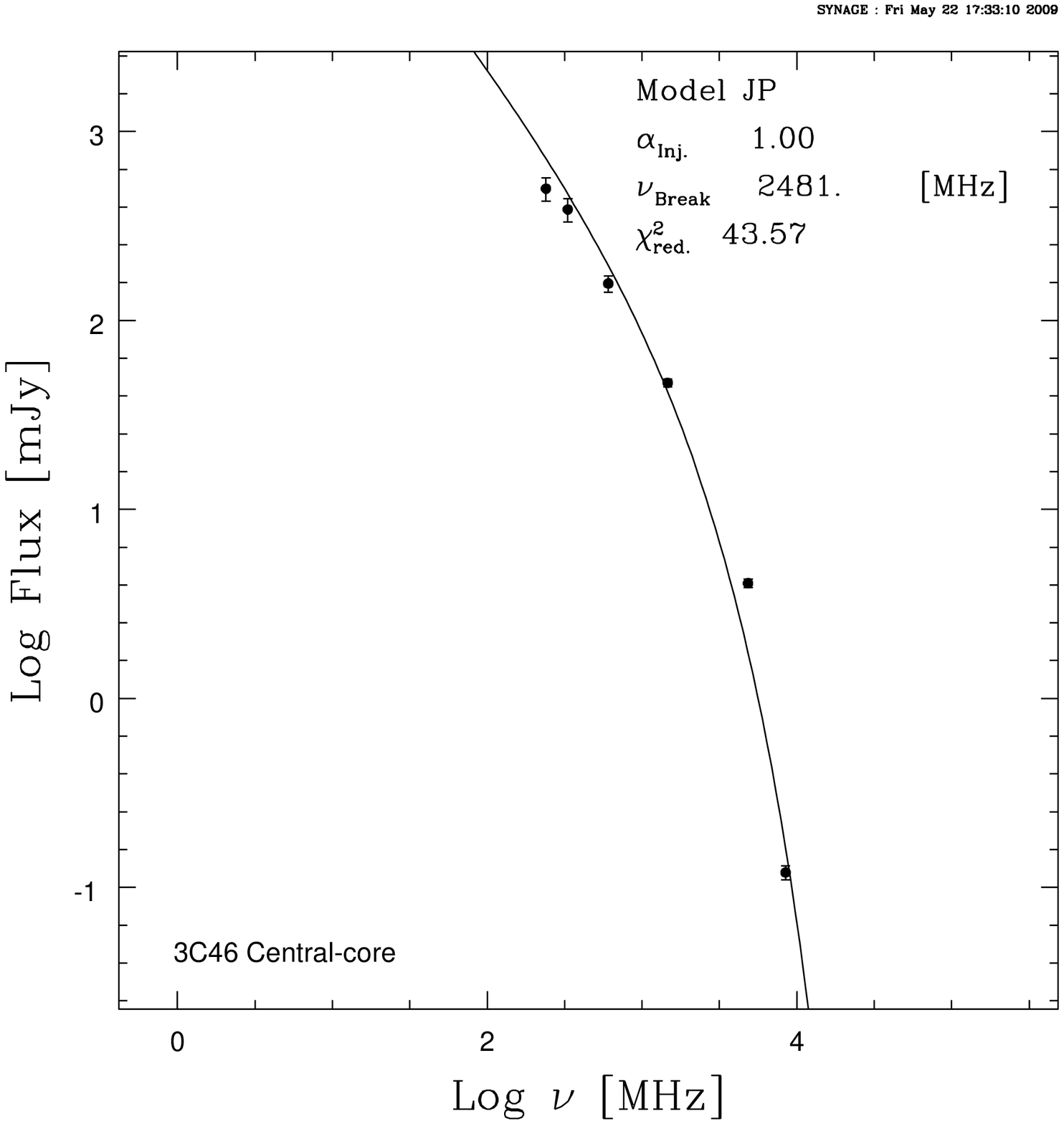,width=3.0in,angle=0}
      \psfig{file=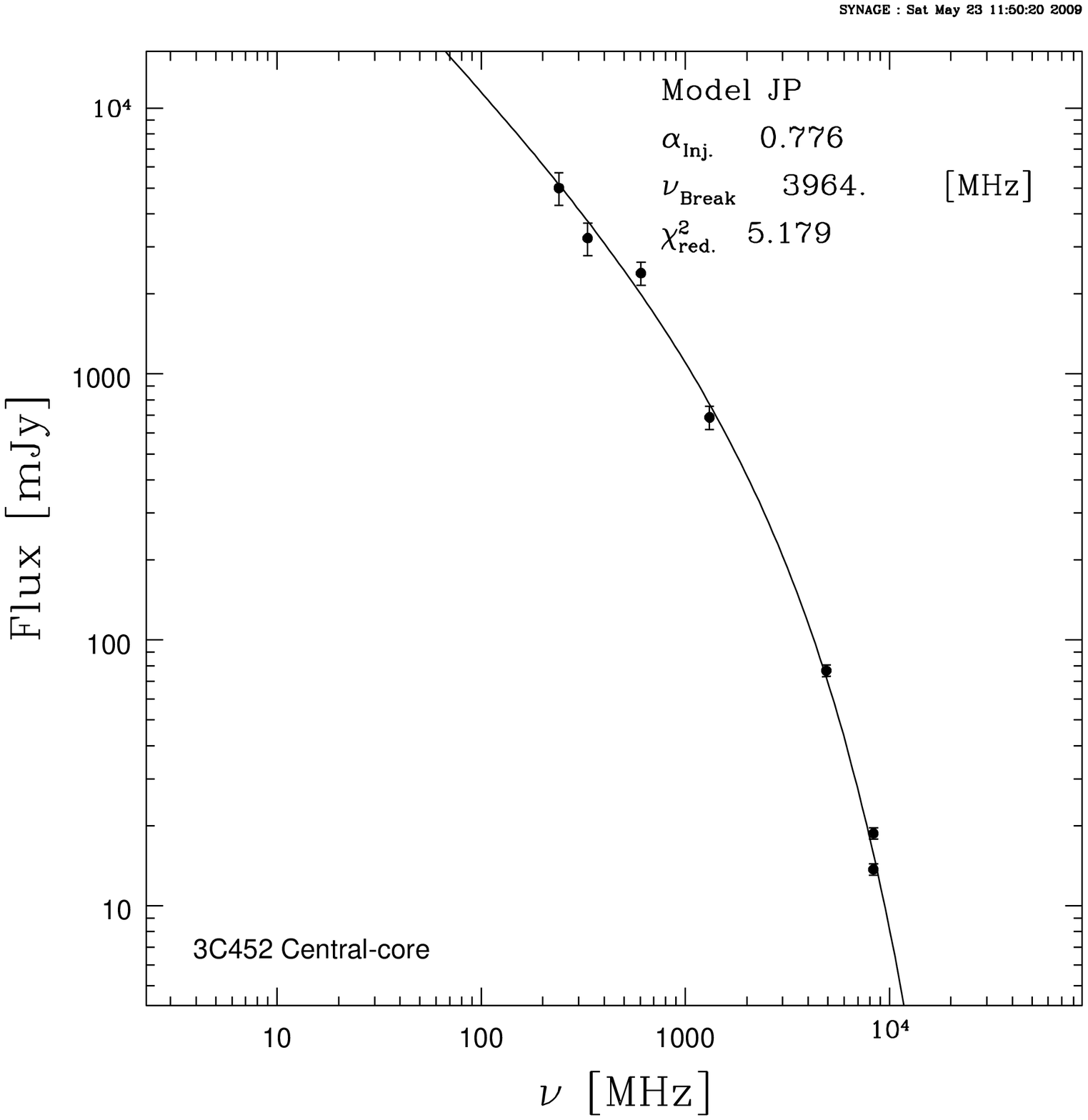,width=3.0in,angle=0}
     }
\caption[]{Spectra of the central regions of 3C46 (upper panel) and 3C452 (lower panel) after subtraction of the
           core flux density, with the fits from the JP model as described in the text. }
\end{figure}
%%%%%%%%%%%%%%%%%%%%%%%%%%%%%%%%%%%%%%%%%%%%%%%%%%%%%%%%%%%%%%%%%%%%%%%%%%%%%%%%%%%%

\begin{table}
\caption {Estimates of break frequency and spectral age for 3C46}
\begin{tabular}{lrcl cl}
\hline
Strip & Dist. & $\nu_{\rm br}$ & $\chi^{2}_{\rm red}$ & B  & $\tau_{\rm spec}$ \\
      & (kpc) &(GHz) &    & nT & (Myr) \\
\hline
      &       &    {\bf Western lobe}               &      &  $\rm\alpha_{inj}=1.0$  &                \\
 W1   & 316   &     $ >100                   $      & 5.42 & 1.64   & $ <1.7  $ \\
 W2   & 237   &     $ >100                   $      & 1.78 & 1.35   & $ <2.2  $ \\
 W3   & 158   &     $36.4^{+6.5}_{-28.3}     $      & 1.71 & 1.35   & $ 3.6^{+4.1}_{-0.4}  $ \\
 W4   &  79   &     $10.6^{+0.7}_{-3.9}      $      & 0.34 & 1.34   & $ 6.7^{+1.7}_{-0.2}    $ \\
      &       &                                     &      &            &   \\
      &       &    {\bf Eastern lobe}               &      &  $\rm\alpha_{inj}=1.0$   &                \\
 E1   & 519   &     $ >100                   $      & 10.2 & 1.50   & $<1.9 $ \\
 E2   & 440   &     $ >100                   $      & 11.4 & 1.36   & $<2.2 $ \\
 E3   & 361   &     $ >100                   $      & 5.1  & 1.20   & $<2.4 $ \\
 E4   & 282   &     $ >100                   $      & 1.0  & 1.12   & $<2.6 $ \\
 E5   & 203   &     $36.3^{>+6.6}_{-25.8}    $      & 0.9  & 1.26   & $ 3.9^{+3.3}_{-0.3}  $ \\
 E6   & 124   &     $9.7^{>+0.6}_{-2.2}      $      & 0.2  & 1.30   & $ 7.3^{+1.0}_{-0.2}  $ \\
 E7   &  56   &     $7.0^{>+0.4}_{-0.3}      $      & 4.1  & 1.22   & $ 9.1^{+0.2}_{-0.3}  $ \\
      &       &                                     &                      &                      \\
      &       &    {\bf Central-core}               &      &$\rm\alpha_{inj}=1.0$ &                \\
      &       &     $2.5^{+0.05}_{-0.7}      $      & 44   & 1.28   & $ 14.5^{+2.6}_{-0.1} $ \\
\hline
\end{tabular}
\end{table}
%%%%%%%%%%%%%%%%%%%%%%%%%%%%%%%%%%%%%%%%%%%%%%%%%%%%%%%%%%%%%%%%%%%%%%%%%%%%%%%%%%%%%%%%%%%%%%%%%%

\begin{table}
\caption {Estimates of break frequency and spectral age for 3C452}
\begin{tabular}{lrcl cl}
\hline
Strip & Dist. & $\nu_{\rm br}$                      & $\chi^{2}_{\rm red}$ & B &     $\tau_{\rm spec}$      \\
      &(kpc)  &(GHz)                                &                      & nT     & (Myr)                \\
\hline
      &       &    {\bf Western lobe}               &        &         $\rm\alpha_{inj}=0.776$ &       \\
W1    & 197     &         $>100$                      & 9.87 & 0.76    &  $<5.9  $                        \\
W2    & 173     &         $>100$                      & 0.71 & 0.83    &  $<5.4  $                         \\
W3    & 150     &         $>100$                      & 0.71 & 0.79    &  $<5.6  $                         \\
W4    & 126     &     $68.8^{+24.7}_{-36.1} $         & 0.49 & 0.77    & $  7.0^{+3.2}_{-1.0}        $ \\
W5    & 103     &     $27.9^{+28.3}_{-11.7} $         & 0.18 & 0.80    & $ 10.6^{+3.3}_{-3.1}        $ \\
W6    &  79     &     $13.6^{+3.4}_{-5.3}   $         & 1.12 & 0.81    & $ 14.9^{+4.2}_{-1.6}        $ \\
W7    &  56     &     $ 7.7^{+0.5}_{-3.8}   $         & 3.71 & 0.82    & $ 19.6^{+7.9}_{-0.6}        $ \\
W8    &  32     &     $ 6.4^{+0.8}_{-1.4}   $         & 8.81 & 0.80    & $ 22.1^{+2.9}_{-1.3}        $ \\
      &         &                                     &      &         &                                  \\
      &         &    {\bf Eastern lobe}               &      &          $\rm\alpha_{inj}=0.776$ &      \\
E1    & 194     &         $>100$                      & 8.30 & 0.76    &   $<5.9   $                       \\
E2    & 171     &         $>100$                      & 2.64 & 0.80    &   $<5.6   $                       \\
E3    & 147     &     $ 60.2^{+34.2}_{-48.8}     $    & 3.13 & 0.82    & $  6.9^{+9.0}_{-1.4}       $ \\
E4    & 123     &     $ 25.6^{+4.8}_{-18.9}      $    & 5.47 & 0.83    & $ 10.5^{+10.0}_{-0.9}       $ \\
E5    & 100     &     $ 17.3^{+4.0}_{-8.6}       $    & 1.88 & 0.83    & $ 12.9^{+5.3}_{-1.3}        $ \\
E6    &  76     &     $ 13.1^{+5.2}_{-3.6}       $    & 0.22 & 0.84    & $ 14.5^{+2.5}_{-2.3}        $ \\
E7    &  53     &     $  9.8^{+2.3}_{-2.07}      $    & 0.53 & 0.83    & $ 17.1^{+2.2}_{-1.7}        $ \\
E8    &  29     &     $ 10.8^{+35.0}_{-0.9}      $    & 0.88 & 0.79    & $ 17.3^{+0.7}_{-8.9}       $ \\
      &         &                                     &      &         &                                  \\
      &         &    {\bf Central-core}               &      &    $\rm\alpha_{inj}=0.776$&                \\
      &         &     $ 3.96^{+0.2}_{-0.8} $          & 5.18 & 0.83    & $ 26.8^{+3.2}_{-0.6}       $ \\
\hline
\end{tabular}
\end{table}

%%%%%%%%%%%%%%%%%%%%%%%%%%%%%%%%%%%%%%%%%%%%%%%%%%%%%%%%%%%%%%%%%%%%%%%%%%%%%%%%%%%%%%%%%%%%%%%%%%%%%%%%%%%%%%%%%%%%%%%%%%%%%%%%%%%%%%

%%%%%%%%%%%%%%%%%%%%%%%%%%%%%%%%%%%%%%%%%%%%%%%%%%%%%%%%%%%%%%%%%%%%%%%%%%%%%%%%%%%%
\begin{figure}
\vbox{\psfig{file=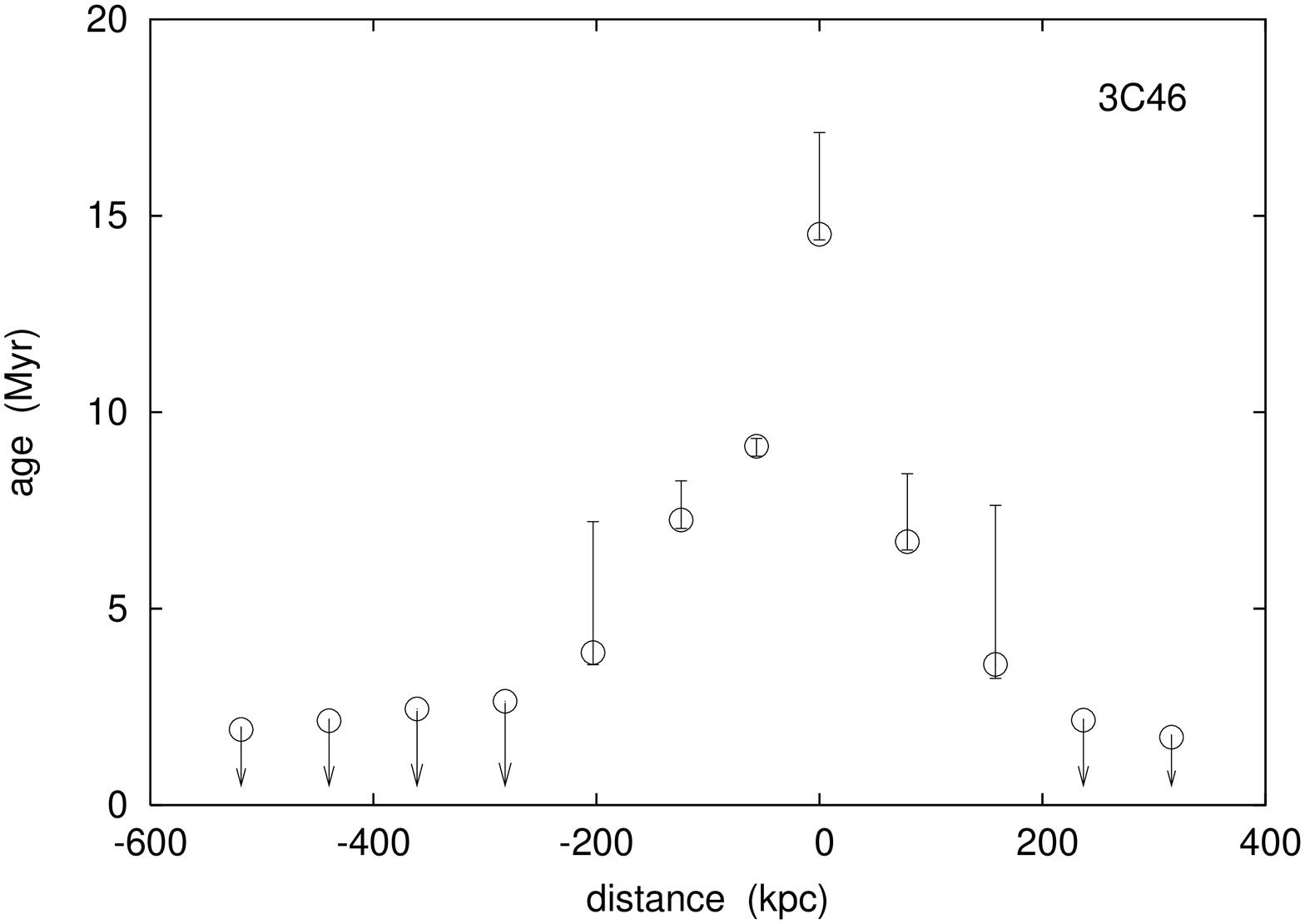,width=3.3in,angle=-90}
      \psfig{file=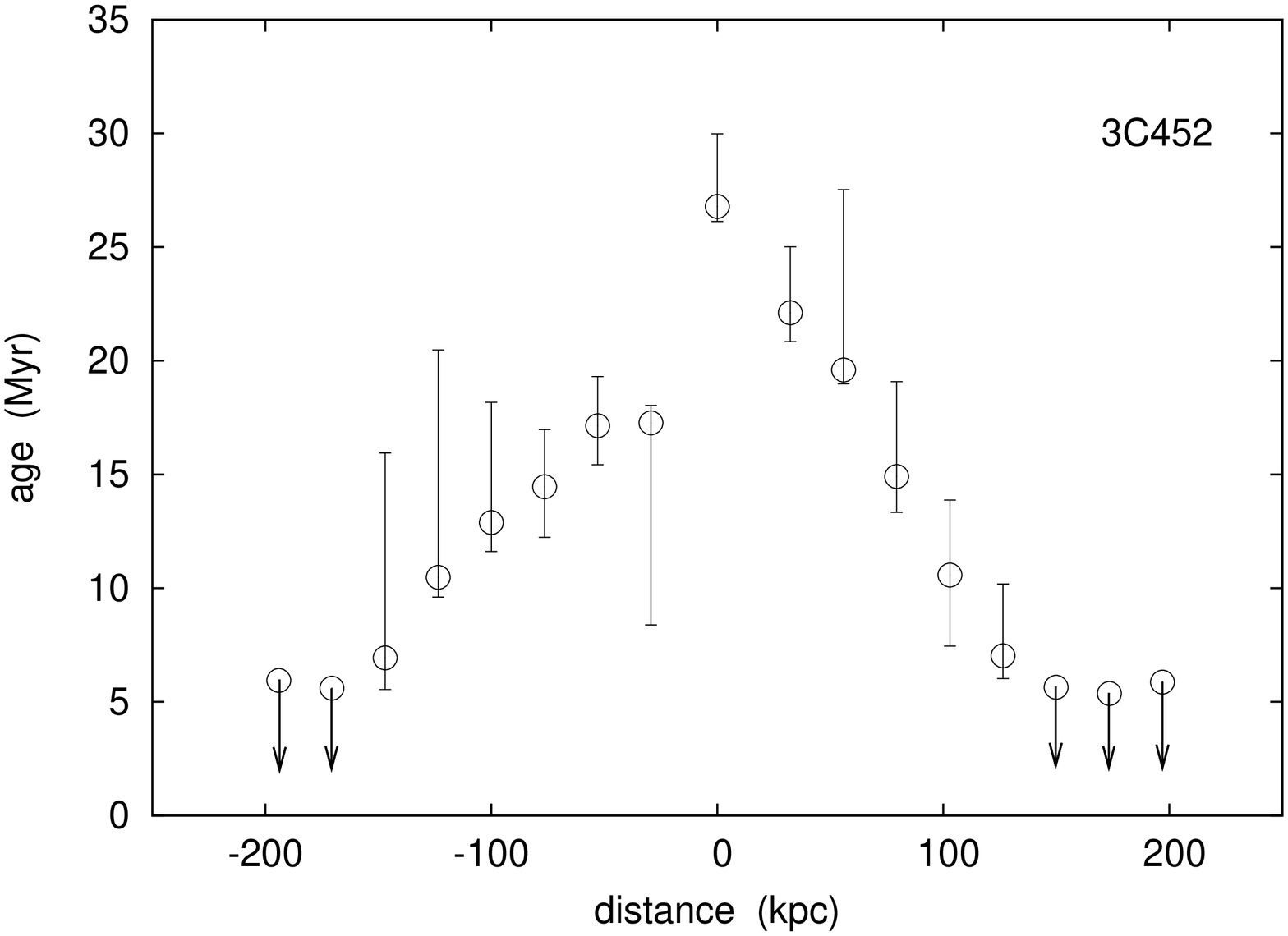,width=3.3in,angle=-90}
     }
\caption[]{The spectral age is plotted against the distance from the core for 3C46 (upper panel) 
           and 3C452 (lower panel) using the magnetic field estimated for each strip. }
\end{figure}
%%%%%%%%%%%%%%%%%%%%%%%%%%%%%%%%%%%%%%%%%%%%%%%%%%%%%%%%%%%%%%%%%%%%%%%%%%%%%%%%%%%%

%%%%%%%%%%%%%%%%%%%%%%%%%%%%%%%%%%%%%%%%%%
In order to estimate the spectral ages, we have to estimate the magnetic-field strength.  
We have { estimated the magnetic field strength by integrating the spectrum from a 
frequency corresponding to a minimum Lorentz factor, $\gamma_{\rm min}\sim$10 for the
relativistic electrons to an upper limit of 100 GHz, which corresponds to a Lorentz 
factor ranging from a few times 10$^4$ to 10$^5$ depending on the estimated magnetic
field strength (see Hardcastle et al. 2004; Croston et al. 2005; Konar et al. 2008, 2009).
It has also been assumed that the filling factors of the lobes are unity,  and the
energetically dominant particles are the radiating particles only, neglecting the 
contribution of the protons. We have assumed a cylindrical geometry for the entire lobe,
and have estimated the magnetic field for the entire lobe as well as for individual 
strips of emission each separated by approximately a beamwidth along the long axis of each source.
The magnetic field strengths for each strip
are listed in Tables 3 and 4 for 3C46 and 3C452 respectively.  
The magnetic field strengths for the western and eastern lobes of 3C46 are 1.66 and 1.53 nT
while the corresponding values for 3C452 are 0.86 and 0.88 nT respectively.}
The equipartition
magnetic fields are usually within a factor of 2 of those estimated from inverse-Compton
scattering of the radiating electrons by the microwave background radiation (e.g. Croston
et al. 2005; Konar et al. 2009). For the giant radio source 3C457, the spectral ages 
estimated from the equipartition magnetic field and the field estimated from the 
inverse-Compton scattered X-ray emission are { similar} (Konar et al. 2009).

The results of our spectral ageing analysis for 3C46 and 3C452 are tabulated in 
Tables 3 and 4 respectively, which are arranged as follows. 
Column 1: identification of the strip,
column 2: projected distance of the centre of the strip from the radio core in units of kpc,
column 3: break frequency of the spectrum of the strip according to the JP model in units of GHz,
column 4: reduced $\chi^{2}$ value of the fit,
column 5: magnetic-field strength in units of nT, and
column 6: spectral age of particles in the strip.
The strips close to the hotspots are consistent with having straight spectra and the 
{\tt SYNAGE} fits yield spectral breaks at frequencies larger than several hundred GHz.
Our observations show no spectral break till $\sim$10 GHz. In Tables 3 and 4 we have listed
the values corresponding to a break frequency of 100 GHz for these strips. Observations
at mm wavelengths are required to determine reliably the spectral breaks in these regions.
{ The spectral age increases with distance from the hotspot for both sources, with the
maximum spectral ages estimated for 3C46 and 3C452 being $\sim$15 and 27 Myr in the
regions closest to the core (Fig. 8).}

\subsection{Search for episodic activity}
The outer diffuse lobes from an earlier cycle of activity in sources with episodic activity
are expected to have a steep spectra due to radiative and adiabatic losses. This has been
observationally demonstrated in some cases such as J1453+3308 (Konar et al. 2006) and
4C29.30 (Jamrozy et al. 2008). Therefore, { ideally} one should be able to detect these 
features more easily at low frequencies. Our low-frequency images show the prominent bridges 
of emission but no diffuse features that could be attributed to an earlier cycle of activity.
{ In the case of 3C46, a diffuse component of say $\sim$20 mJy at 153 MHz and a spectral index
of 1 would have a surface brightness similar to the rms noise in the 8460-MHz image. The 
corresponding value for 3C452 is $\sim$8 mJy at 153 MHz. These values would increase by a factor of $\sim$2
and 7 for spectral indices of 1.2 and 1.5 respectively. With the rms noise values of 
6.9 and 8.7 mJy beam$^{-1}$ at 153 MHz for 3C46 and 3C452 respectively, diffuse emission 
not seen at the highest frequency could have been just about detected in 3C46. However, for 3C452 only
steeper spectrum emission with a spectral index of $\sim$1.4 could have been detected.
While it is important to make more sensitive images at the lowest frequencies, the non-detection
of extended emission due to an earlier cycle of activity in these two sources, is} 
consistent with the trend that such objects are rare even amongst large radio
sources (cf. Schoenmakers et al. 2000; Saikia et al. 2006).

\section{Concluding remarks}
The maximum spectral ages determined for 3C46 and 3C452 are { $\sim$15 and 27 Myr respectively}, 
which is similar to the values of Jamrozy et al. (2008) obtained for a sample of 10 large radio
galaxies by combining GMRT and VLA data. Their values range from $\sim$6 to 36 Myr with 
a median value of $\sim$20 Myr using the classic equipartition magnetic fields. 
These estimates are significantly older than those of smaller sources 
(e.g. Leahy et al. 1989;  Liu et al. 1992), and broadly consistent with the tendency
for spectral age to increase with the projected linear size (Jamrozy et al. 2008 and 
references therein).

The injection spectral indices are $\sim$1.0 and 0.78, compared with the values ranging
from 0.55 to 0.88 with a median value of $\sim$0.6 for the sample of Jamrozy et al. (2008). 
Our estimates for 3C46 and 3C452, which have prominent hotspots, are consistent with the
higher values in the sample of Jamrozy et al. (2008), and  studies of smaller 
FRII sources studied by Leahy et al. (1989) and  Liu et al. (1992). Our estimates of the
injection spectral indices appear steeper than theoretically expected values for 
a strong, non-relativistic shock in a Newtonian fluid where $\alpha_{\rm inj}$ = 0.5 
(Bell 1978a,b; Blandford \& Ostriker 1978), or for different scenarios involving 
relativistic shocks where  $\alpha_{\rm inj}$ vary in the range of 0.35
to 0.65 (Heavens 1989; Kirk \& Schneider 1987; Drury \& Volk 1981; Axford, Leer
\& McKenzie 1982). High-resolution observations of these sources at even lower frequencies
with future instruments should help in determining injection spectra more reliably.

%%%%%%%%%%%%%%%%%%%%%%%%%%%%%%%%%%%%%%%%%%%%%%%%%%%%%%%%%%%%%%%%%%%%%%%%%%%%%%%%%%%%%
\section*{Acknowledgments}
We thank an anonymous reviewer for a very detailed report which has improved the
paper significantly.
SN and AP thank NCRA, TIFR for hospitality during the course of this work and DST,
Government of India for financial support vide Grant No. SR/S2/HEP-17/2005.  We thank 
Neeraj Gupta for providing the calibrated L-band GMRT data on 
3C452. The GMRT is a national facility operated by the National Centre 
for Radio Astrophysics of the Tata Institute of Fundamental Research. We thank the staff
for help with the observations.  The National Radio Astronomy Observatory  is a
facility of the National Science Foundation operated under co-operative
agreement by Associated Universities Inc. We thank the VLA staff for easy access
to the archival data base.  This research has made use of the NASA/IPAC extragalactic database (NED)
which is operated by the Jet Propulsion Laboratory, Caltech, under contract
with the National Aeronautics and Space Administration.  We thank numerous contributors
to the GNU/Linux group.

%%%%%%%%%%%%%%%%%%%%%%%%%%%%%%%%%%%%%%%%%%%%%%%%%%%%%%%%%%%%%%%%%%%%%%%%%%%%%%%%%%%%%
{}

\end{document}